\documentclass[acmlarge,screen]{acmart}

\AtBeginDocument{%
  }

\setcopyright{cc}
\setcctype{by-nc-nd}
\acmJournal{IMWUT}
\acmYear{2026} \acmVolume{10} \acmNumber{3} \acmArticle{194}
\acmMonth{9} \acmDOI{10.1145/3831965}

\usepackage{xcolor}
\usepackage[normalem]{ulem}
\usepackage{booktabs}
\usepackage{pifont}

\newif\ifshowrevisions
\showrevisionsfalse   

\ifshowrevisions
  \newcommand{\rev}[1]{\textcolor{blue}{#1}}
  \newcommand{\del}[1]{\textcolor{red}{\sout{#1}}}
\else
  \newcommand{\rev}[1]{#1}
  \newcommand{\del}[1]{}
  
\fi

\usepackage{multirow} 
\begin{document}

\title{Mammal: Supporting Breastfeeding Monitoring Through Computational Garments with Inter-Body Sensing}

\author{Yanfeng Zhao}
\email{yz24f@fsu.edu}
\orcid{0009-0008-0668-0429}
\affiliation{%
  \institution{Florida State University}
  \city{Tallahassee}
  \state{Florida}
  \country{USA}
}

\author{Morgan Geck}
\email{mlgeck@ncsu.edu}
\orcid{0009-0008-3675-1822}
\affiliation{%
  \institution{North Carolina State University}
  \city{Raleigh}
  \state{North Carolina}
  \country{USA}
}

\author{Kate Fernandez}
\email{kf21g@fsu.edu}
\orcid{0009-0005-3049-8537}
\affiliation{%
  \institution{Florida State University}
  \city{Tallahassee}
  \state{Florida}
  \country{USA}
}

\author{Madison Nicole Jones}
\email{mnj22d@fsu.edu}
\orcid{0009-0008-2780-6940}
\affiliation{%
  \institution{Florida State University}
  \city{Tallahassee}
  \state{Florida}
  \country{USA}
}

\author{Xia Zhou}
\email{xia@cs.columbia.edu}
\orcid{0000-0002-2852-9024}
\affiliation{%
  \institution{Columbia University}
  \city{New York}
  \state{New York}
  \country{USA}
}

\author{Jessica L. Ridgway}
\email{jridgway@fsu.edu}
\orcid{0000-0001-7632-6160}
\affiliation{%
  \institution{Florida State University}
  \city{Tallahassee}
  \state{Florida}
  \country{USA}
}

\author{Te-Yen Wu}
\email{teyen.wu@fsu.edu}
\authornote{Corresponding author.}
\orcid{0000-0003-3977-9093}
\affiliation{%
  \institution{Florida State University}
  \city{Tallahassee}
  \state{Florida}
  \country{USA}
}

\renewcommand{\shortauthors}{Zhao et al.}


\begin{abstract}
Breastfeeding provides critical insight into infant feeding competence and physiological health, yet objective monitoring remains difficult due to the intimate and internal nature of feeding. We present Mammal, a caregiver-worn computational garment that unobtrusively monitors breastfeeding without attaching sensors to the infant. Mammal leverages inter-body signal transmission through natural mouth-to-breast contact to capture infant cardiac and feeding-related acoustic signals on the caregiver’s body.
Using novel algorithms to detect latch onset, \rev{infer} infant electrocardiogram (ECG), and identify suck and swallow events from inter-body signals, Mammal estimates latch duration, in-feeding heart rate, suck–swallow–breathe (SSB) ratio, and milk intake.
In a user study with \rev{10} caregiver–infant dyads, Mammal achieves a mean absolute percentage error (MAPE) of \rev{5.56\%} for latch duration, a mean absolute error (MAE) of \rev{3.61} bpm for infant heart rate estimation, a mean absolute error of \rev{0.12} for SSB ratio estimation, and a mean relative error of \rev{15.76\%} for milk intake, with participants reporting high comfort and wearability.
\end{abstract}


\begin{teaserfigure}
  \includegraphics[width=\textwidth]{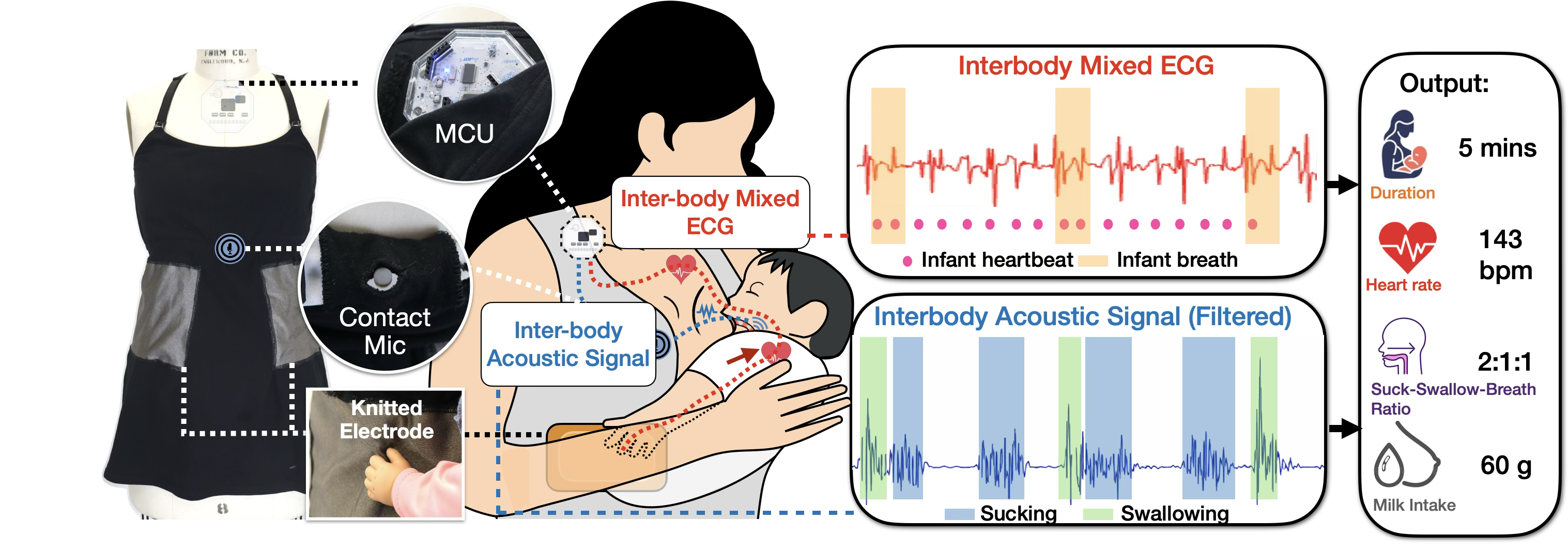}
  \caption{We introduce Mammal, a caregiver-worn computational garment that monitors daily breastfeeding using inter-body sensing to derive key indicators, including latch duration, heart rate, suck–swallow–breathe ratio, and milk intake.}
  \Description{}
  \label{fig:teaser}
\end{teaserfigure}


\maketitle

\section{INTRODUCTION}
Breastfeeding is one of the most common and essential baby care activities, widely recommended as the optimal method for infant feeding~\cite{Victora01,haruka}. Beyond its nutritional benefits, it offers valuable insights into infant health. Breastfeeding reflects an infant’s ability to coordinate oral structures and muscle control~\cite{WHO2013, Wen-jen08}. Difficulties in oral–motor coordination and neuromuscular development can impair breastfeeding~\cite{Muro09, Lau11}, leading to irregular sucking patterns and disruptions in the suck–swallow–breathe rhythm~\cite{Lau12, Gewolb14}. Moreover, the cardiovascular effort required during feeding can expose hidden heart conditions such as bradycardia or tachycardia~\cite{suiter_oxygen_2007, veerappan_spectral_2000, jcdd12020038}. \rev{Monitoring breastfeeding therefore not only enables understanding of infant nutritional intake, but also supports early detection of feeding challenges and provides a non-invasive window into an infant’s neurological, muscular, and cardiovascular health.}

Despite its importance, monitoring breastfeeding is difficult. Unlike bottle feeding, where intake volume, feeding duration, and even oral activities can be easily measured~\cite{goldfield2006coordination,7577837}, breastfeeding is a largely internal and intimate process that offers limited visibility into what occurs during feeding. \rev{Although approaches such as pre- and post-feed weighing or in-hospital equipment, including wired electrocardiogram sensors and respiratory belts, can provide partial information, they are often cumbersome, intrusive, and impractical for routine use, especially in home settings where most breastfeeding takes place. These methods may also require repeated handling, specialized equipment, or skin-contact sensors that can disrupt natural feeding and irritate an infant’s sensitive skin. As a result, there is an urgent need for non-intrusive breastfeeding monitoring solutions that can be used comfortably at home~\cite{kim2025compact,Kymeng04}. Such tools would enable early identification of feeding issues, support caregiver confidence, and promote sustained breastfeeding practices outside clinical settings.}


In this paper, we present Mammal, a wearable computational garment designed for caregivers to unobtrusively monitor breastfeeding sessions and infants’ physiological states (Fig.~\ref{fig:teaser}). Mammal measures key indicators such as latch duration, infant heart rate (HR), the suck–swallow–breathe ratio, and milk intake. The core principle of Mammal lies in inter-body signal transmission. 
Infants' physiological signals such as ECG and acoustic signals from sucking and swallowing are transmitted through the natural mouth-to-breast contact and collected by Mammal without requiring direct attachment of wires or probes to the infant. While these signals are mixed with the caregiver’s own bio-signals, they can be separated and analyzed using advanced signal processing techniques to derive measurements of the infant’s physiological activities. This approach not only preserves the intimacy and comfort of breastfeeding but also enables non-invasive monitoring that provides caregivers with insights into infant health and feeding behaviors.

However, realizing this computational garment faces three challenges. First, identifying latching is crucial for understanding feeding duration and behaviors. However, due to the subtle nature of this interaction, it is difficult to determine when the infant’s mouth makes contact with the breast to initiate feeding. This goes beyond the latest skin-to-skin monitoring using inter-body ECG signals~\cite{shao2024joey}, where physical contact is not confined to mouth and breast. Second, while prior research has demonstrated the detection of sucking and swallowing by placing sensors near the mouth and throat~\cite{fukuike, yatani2012bodyscope}, a critical knowledge gap remains in sensing these events through inter-body acoustic signals. When acoustic signals propagate across the mouth-to-breast contact, they are attenuated, raising questions about how detection can be achieved in a caregiver-worn system. Lastly, capturing inter-body ECG to derive infant's heart rate requires the infant’s body to maintain contact with a textile electrode on the garment to establish an electrical conduction path across two bodies. Achieving this across diverse breastfeeding postures and contexts necessitates an ergonomically designed garment that ensures consistent functionality while preserving comfort.

To address these challenges, we first examine the principles of inter-body transmission. We conduct experiments to understand how electrical and acoustic signals may transmit through the coupled bodies of the caregiver and infant during breastfeeding. We observe that mouth-to-skin (breast) contact exhibits stronger electrical coupling compared to general skin-to-skin contact, enabling Mammal to detect latch events by measuring coupling strength. In parallel, we characterize inter-body acoustic signals generated by sucking and swallowing and identify the frequency bands that remain observable after attenuation. Guided by these findings, we develop a hardware–software co-design. The garment’s electrode and sensor layout are optimized for common breastfeeding postures and practices. A signal processing pipeline is developed to \rev{infer infant's ECG}, detect latch, sucking and swallowing events from inter-body signals. This enables the estimation of latch duration, in-feeding heart rate, suck–swallow–breathe ratio, and milk intake.



To \rev{understand} the performance of Mammal, we fabricate a tank-top–style prototype in collaboration with fashion designers and conduct a user study with \rev{10} caregiver–infant dyads. 
Overall, Mammal achieves a mean absolute percentage error of \rev{5.56\%} for latch duration estimation, a mean absolute error of \rev{3.61} beats per minute for infant heart rate estimation, a mean absolute error of \rev{0.12} for SSB ratio estimation, and a \rev{15.76\%} mean relative error for breastmilk intake estimation. Participants reported that Mammal is comfortable, easy to wear, and compatible with daily breastfeeding routines. In addition, a preliminary interview with a neonatal intensive care unit (NICU) clinician suggested that Mammal could provide practical and clinically meaningful support during the NICU-to-home transition. Finally, we discuss \rev{several limitations} and key next steps, including a longitudinal validation study and hardware improvement for washability and maintenance.

The contributions of this work are as follows:
\begin{itemize}
    \item We characterize the inter-body transmission mechanisms of electrical and acoustic signals to understand how physiological signals transmit across mouth-to-skin (breast) contact.

    \item We design and fabricate an ergonomic tank-style nursing garment that integrates textile electrodes and contact microphones to enable comfortable and unobtrusive signal acquisition across diverse breastfeeding scenarios.

    \item We develop signal processing algorithms to \rev{infer infant’s ECG, and identify latch, sucking, and swallowing events from inter-body electrical and acoustic signals, enabling the derivation of key feeding indicators such as latch duration and suck–swallow–breathe ratio.} 

    \item We \rev{evaluate} the performance of Mammal in a user study with \rev{10} caregiver–infant dyads, demonstrating accurate detection of latch onset and duration, robust inference of physiological features, and high user comfort and usability.
\end{itemize}

\section{RELATED WORK}

\subsection{Physiological Monitoring for Infants}

Infant physiological monitoring has been widely studied in both clinical and home contexts, with a primary focus on vital signals such as respiration, heart rate, and heart rate variability~\cite{sullivan2022vital}. In clinical settings, these measurements are traditionally obtained using wired and stationary equipment deployed at scale in neonatal care environments~\cite{sullivan2021clinical}. While such systems provide reliable signal quality, they require direct attachment to the infant and are typically confined to supervised clinical use. More recently, wireless and wearable sensing solutions have been introduced into clinical and home settings, including sensing socks~\cite{malik2020owlet,fda2023den220091dreamsock}, ankle bands~\cite{savage2020sibel}, instrumented sleep surfaces~\cite{liuchen}, and electronic skin~\cite{chung2020skin}. These sensing systems monitor infants' vital signs but do not capture feeding-related behaviors. Therefore, a complementary line of work explores physiological sensing for feeding. Prior studies have instrumented objects, such as feeding bottles with pressure sensors, to assess sucking strength and diagnose feeding deficits~\cite{chen2018bdd}. Yet, these approaches primarily focus on bottle-feeding and do not leverage \rev{inter-body sensing} for monitoring breastfeeding physiology.

\rev{Joey is the first system to capture inter-body ECG signals during skin-to-skin care for inferring infants' heart rate and respiration rate~\cite{shao2024joey}. Building on this line of inter-body sensing, Mammal targets breastfeeding, which requires a garment form factor that supports breastfeeding postures and sustained caregiver--infant contact. Mammal further extends inter-body ECG from physiological monitoring to latch detection by estimating caregiver--infant ECG coupling strength to distinguish contact types (e.g., skin-to-skin vs.\ mouth-to-skin). To operate in breastfeeding contexts, Mammal also introduces an ECG separation algorithm robust to feeding-related sucking artifacts, which introduce non-stationary structured noise not addressed in Joey. Finally, Mammal incorporates inter-body acoustic sensing to capture sucking and swallowing through the caregiver's body, enabling breastmilk intake estimation and multimodal inference of SSB patterns.}


\del{In summary, while existing work establishes the feasibility of wearable infant monitoring, a gap remains in capturing the multimodal dynamics of natural feeding routines. Most prior systems focus either on isolated vital signs or on object-dependent feeding metrics. Our work builds on this foundation by introducing a caregiver-worn, multimodal sensing garment that captures breastfeeding coordination, vital signs, and feeding duration, enabling session-level analysis of infant physiological dynamics during breastfeeding.}

\subsection{Body Signal Transmission}
Human body as a signal channel for interaction sensing or device-to-device communication has been widely explored in HCI and mobile computing.
Prior work mainly falls into two families: \emph{electrical} and \emph{acoustic} channels.
In the electrical family, \rev{body-coupled communication (BCC) or human body communication (HBC)}  leverages the human body as a conductive path to carry modulated signals, enabling touch sensing~\cite{ActiTouch,SkinTrack}, gesture sensing~\cite{Tomo}, and hand-held object sensing~\cite{EMsense}.
It has also been used as an enabling primitive for interactive infrastructure and on-/near-body device communication, where devices exchange data only when the user physically touches an object or surface~\cite{TouchCom,VargaUISTGroundlessBCC,BodyWireHCI}.
On the other hand, the acoustic family treats the human body as a vibro-mechanical waveguide, where taps, motion, and internal physiological events generate vibrations that propagate through tissue and bone.
This channel has been widely used for on-body input and touch localization~\cite{Skinput}, activity and context sensing~\cite{Viband}, health sensing~\cite{goverdovsky_hearables_2017} and device communication~\cite{Vricomm}.

However, the majority of prior work in both electrical and acoustic families focuses on \emph{intra-body} signal transmission, where sensing or communication occurs within a single user’s body.
Only limited work has begun to explore inter-body channels formed through physical contact or close proximity between two people.
For example, prior studies have characterized inter-body electrical coupling during handshakes or touch events, demonstrating the feasibility of signal transmission across bodies but primarily from a physical-layer or security perspective~\cite{li2017handshake, 8513332}.
The most relevant work, Joey, investigates inter-body electrical transmission under infant-caregiver skin-to-skin contact and establishes signal models for inter-body ECG transmission~\cite{shao2024joey}. Building on this foundation, our work similarly leverages electrical inter-body transmission, but further incorporates acoustic inter-body signals to infer feeding-related events such as sucking and swallowing, enabling a richer characterization of breastfeeding physiology.

\section{SENSING OBJECTIVES AND CHALLENGES}
In this work, we aim to develop a computational garment that enables caregivers to non-invasively monitor key physiological indicators in the breastfeeding process. In the following, we describe each sensing objective and discuss the corresponding technical challenges.

\textbf{Latch Onset and Duration.}
Latch is the period when the infant maintains mouth--breast contact for sucking and milk transfer~\cite{kronborg2009effective}. Identifying when latch begins and how long it is sustained is essential for characterizing feeding behavior and efficiency~\cite{neifert2004breastmilk}. Disrupted latch can reflect maternal factors (e.g., nipple inversion or engorgement~\cite{douglas2022re}) or infant factors (e.g., weak suction~\cite{cadwell2007latching}), often leading to prolonged feeds~\cite{westerfield2018breastfeeding} and reduced transfer efficiency~\cite{kramer2012optimal}. However, measuring latch onset and duration remains challenging due to the intimate nature of breastfeeding. Direct sensing at the nipple is impractical because it may cause discomfort, interfere with natural feeding behavior, and raise hygiene concerns. As a result, we need a new approach that relies on indirect sensing to detect latch.

\textbf{Suck–Swallow–Breathe Ratio.}
SSB coordination reflects neuromuscular maturation and respiratory control. Typical patterns approximate 1:1:1 or 2:1:1 \cite{mizuno2003maturation, bulock1990development}. Persistent deviations (e.g., irregular breathing or disrupted rhythm) may indicate dysphagia or impaired oral--motor/respiratory regulation \cite{dacosta2008sucking, miller2004relationship, sakalidis2012oxygen}. Conditions ranging from structural anomalies (e.g., laryngomalacia \cite{simons2016laryngomalacia}, cleft palate \cite{miller2011feeding}) to rarer disorders (e.g., H-type TEF \cite{wolfe2024tracheoesophageal}, PBVCP \cite{lechien2024management}) can also affect feeding coordination. However, current clinical practices rely on clinician observation and lack non-invasive tools to detect such early physiological warning signs in natural, non-clinical environments. By unobstructively capturing the SSB ratio during breastfeeding, we can thus provide valuable insights into infant development and health.


\textbf{Heart Rate.}
Feeding is the primary form of physical activity in early infancy, during which the infant’s cardiovascular and respiratory systems are actively engaged to support rhythmic sucking, swallowing, and breathing. This increased physical demand can accentuate underlying cardiological conditions, such as congenital heart defects or arrhythmias, revealing abnormalities that might otherwise remain unnoticed at rest \cite{medoffcooper2016association,ross2012classification}. Therefore, continuous heart rate monitoring during breastfeeding is valuable.
However, attaching sensing devices directly to the infant’s body is impractical due to their fragile skin and the need to preserve comfort and natural bonding. Thus, alternative approaches, similar to \cite{shao2024joey}, must be used to infer cardiovascular activity during the feeding.

\textbf{Breastmilk Intake.} \rev{Milk intake is central to assessing feeding adequacy, hydration, and early growth, making it an important measure in pediatric care \cite{ABMProtocol2_2022, AAP_PediatricNutrition_Infancy}. Yet, it remains difficult to measure during breastfeeding because milk transfer occurs internally and cannot be directly observed. This lack of direct measurement limits objective clinical feeding guidelines and can increase caregiver uncertainty about feeding adequacy, a concern associated with lower breastfeeding self-efficacy, greater anxiety, and earlier breastfeeding cessation \cite{Gatti_InsufficientMilk, Whipps_Demirci_2021}. Test weighing is a standard approach, but it is burdensome for routine at-home use, can disrupt natural mother--infant interaction \cite{InfantFeeding_Scanlon}, and has limited precision, with reported errors ranging from 7\% to 40\% depending on adherence to standardized weighing protocols \cite{testweighing_Savenije, BORSCHEL1986367, InfantFeeding_Scanlon}.}
Other direct measurement methods often require invasive instrumentation (e.g., flow sensors \cite{prime2009milkflow} or bottles \cite{chen2018bdd}) that can also disrupt natural feeding. Recent noninvasive approaches using breast electrical impedance~\cite{kim2025compact} still require breast instrumentation that may disrupt breastfeeding activities and be prone to disturbance by the feeding infant. This motivates an alternative intake estimation method that preserves natural breastfeeding behavior while enabling practical and acceptably accurate monitoring for daily use.


\section{MAMMAL SENSING PRINCIPLE}

To address these challenges and enable non-invasive sensing of the target objectives, Mammal is designed to use the caregiver’s body as a sensing interface, grounded in the principle of inter-body signal transmission. Specifically, when the infant latches, mouth-to-breast contact forms two coupled pathways (Fig. 2a):

\textbf{Electrical pathway:} The moist oral seal and compliant soft tissue create a low-impedance conductive pathway between infant oral tissue and the nipple--areolar skin. This pathway allows infant bio-potentials (including ECG) to be conducted onto the caregiver’s body surface. As a result, electrodes that measure differential potential between the infant and caregiver can derive an inter-body ECG channel that contains a mixture of attenuated caregiver ECG and infant ECG component. 

\textbf{Mechanical/acoustic pathway:} When the infant seals and applies suction on the nipple--areolar complex, the nipple, infant oral tissues, and caregiver breast tissue become mechanically coupled, creating a pathway for vibration and pressure fluctuations to propagate across the contact. Sucking can thus generate micro-vibrations at the nipple contact, and swallowing can introduce brief, impulse-like vibration to the caregiver's body when the bolus is transported through the oral--pharyngeal structures. 

Together, these pathways provide complementary observables that support Mammal's sensing objectives (Fig.~2b). First, latch onset can be indicated by the emergence and strengthening of inter-body electrical coupling, which should be distinguishable from incidental skin-to-skin contact due to the low impedance of mouth-to-breast contact. Second, infant heart rate is expected to be measurable, given the infant ECG component lies in the inter-body ECG. Third, SSB ratio can be estimated by inferring breathing-related modulation from infant ECG and detecting suck and swallow from inter-body acoustics, then aligning their timing. Finally, milk intake can be estimated by aggregating per-suck milk transfer from detected sucks using the relationship between milk transfer rate and inter-suck interval \cite{bowenjones1982milkflow}.

\begin{figure}
    \centering
    \includegraphics[width=0.8\linewidth]{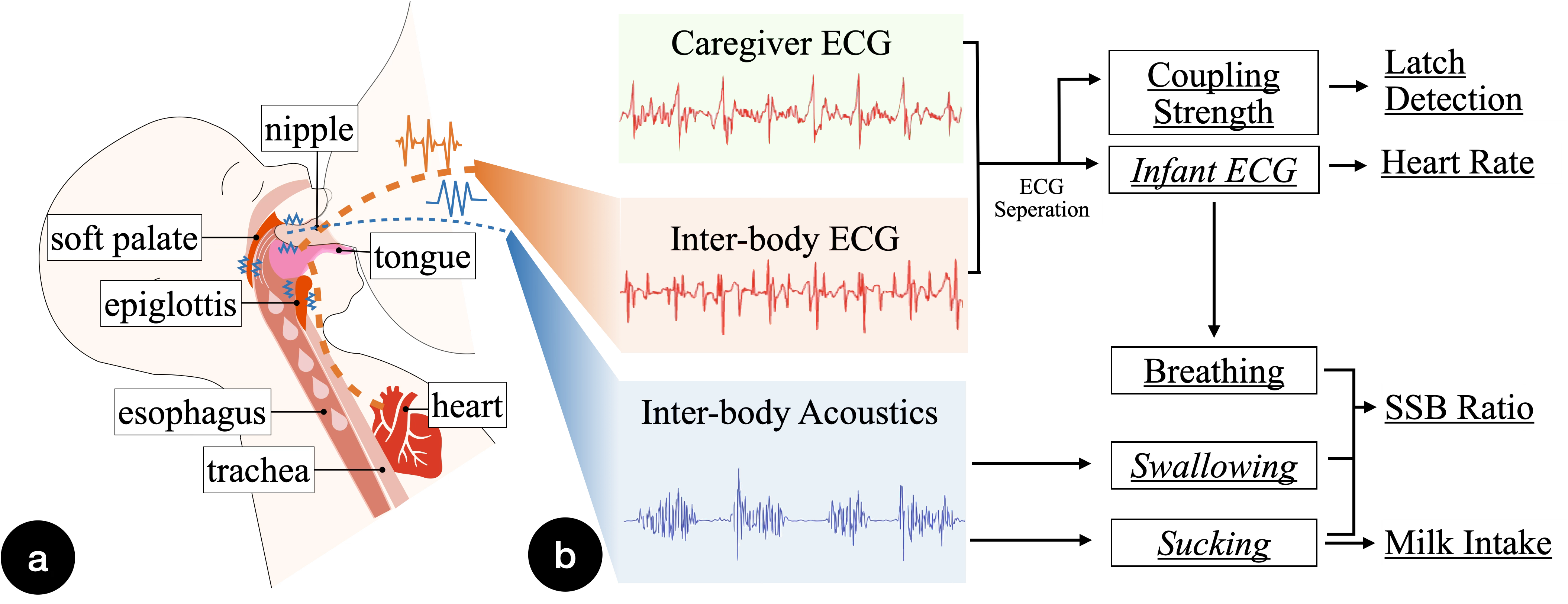}
    \caption{(a) Conceptual illustration of inter-body signal transmission during breastfeeding, showing how infant ECG signals and sucking–swallowing acoustics propagate through the nipple and caregiver’s body. (b) Overview of Mammal’s sensing pipeline, illustrating how these inter-body electrical and acoustic signals can be processed to achieve the sensing objectives.}
    \label{fig:placeholder}
\end{figure}

\section{EXPERIMENTS - INTER-BODY TRANSMISSION CHARACTERIZATION}

Despite the promise of this sensing principle, its feasibility relies on three assumptions that are not yet well-characterized. First, mouth-to-breast contact establishes a distinguishably stronger electrical coupling than general skin-to-skin contact, owing to the lower contact impedance introduced by oral moisture. Second, inter-body ECG signals remain stable despite continuous movements caused by infant sucking during feeding. Third, sucking- and swallowing-induced acoustic signals are able to propagate to the caregiver’s body and remain detectable despite attenuation at the mouth-to-breast contact. To address these assumptions, we conduct two experiments to characterize how electrical and acoustic signals may transmit across the mouth-to-breast contact.


\subsection{Experiment 1 – Inter-body ECG Transmission}

The goal of this study is to investigate how ECG signals transmit across the two human bodies during breastfeeding. We aim to examine the strength and stability of ECG signals transmitted through the mouth-to-breast contact, and compare it with general skin-to-skin transmission. 
However, due to the intimate nature of breastfeeding and the difficulty of maintaining controlled conditions with real caregiver–infant dyads, we instead evaluate inter-body ECG transmission by comparing mouth-to-skin contact with skin-to-skin contact with adult dyads to approximate the scenario.


\textbf{Experiment Setup.} In our study, we recruited 8 adult participants (6 males, 2 females, aged 21-28) and formed 4 dyads. Each dyad of participant was tested under three contact conditions: (1) stable mouth-to-skin contact, where one participant gently touched the lip to the back of the other’s hand; (2) mouth-to-skin contact with rhythmic sucking motions, performed on the same hand location; and (3) general skin-to-skin contact, where participants held hands. Each condition was held for 5 minutes.  This experimental design allows comparative analysis of signal strength and stability across conditions relevant to real breastfeeding interactions.

To collect inter-body and individual ECG signals, we used the OpenBCI Cyton Biosensing Board~\cite{Cyton}. For inter-body ECG, a ground and positive electrode, both made from conductive textile (Adafruit, Knit Conductive Fabric \cite{adafruit1167}), were attached to one participant’s hand, representing the caregiver. The negative electrode, also textile-based, was held by the other participant to simulate the infant. This configuration establishes an inter-body electrical pathway, allowing us to measure how ECG signals transmit between bodies under different contact conditions. For individual ECG, we recorded each participant’s clean, unmixed cardiac signal using a bipolar limb-lead I (Lead I) configuration, with two adhesive electrodes (Kendal, ECG Electrodes \cite{openbci_electrodes}) placed on the right and left upper arm and a third electrode serving as ground \rev{\cite{10.1145/3478127,doi:10.1161/CIRCULATIONAHA.108.191095}}. \rev{
We selected this configuration instead of a 3-lead chest placement because our goal was not full diagnostic ECG acquisition, but capture of QRS-dominant cardiac timing features for downstream heart-rate and respiration-rate inference~\cite{roberts2024opensource, varon2020comparative, earsleeve_liu}.}




\textbf{Data Analysis.} To analyze the collected data, we first \rev{applied a 4th-order Butterworth band-pass filter (2--37~Hz) to all ECG signals, removing baseline drift and high-frequency noise.} To examine whether mouth-to-skin contact exhibits stronger electrical coupling than general skin-to-skin contact, we quantify inter-body contact impedance using ECG signal characteristics. Specifically, we detected peaks from the inter-body ECG signal as well as from each individual ECG signal. These peaks correspond to the \emph{R peaks}, which represent ventricular depolarization and are the most prominent features in an ECG waveform. In theory, R-peak amplitudes observed in the inter-body ECG are attenuated relative to those in individual ECGs because electrical transmission between bodies occurs through a contact interface that introduces additional impedance. To quantify the strength of this coupling, we computed the ratio between the R-peak amplitude measured in an individual ECG and that measured in the mixed inter-body ECG. This \emph{R-amplitude ratio} serves as a proxy measure of inter-body electrical coupling strength and is defined as:
\[
\text{R-Amplitude Ratio}(t)=\frac{A_{\mathrm{R,inter-body}}(t)}{A_{\mathrm{R,individual}}(t)}.
\]

where $A_{\mathrm{R,individual}}(t)$ denotes the R-peak amplitude from either individual’s ECG at time $t$, and $A_{\mathrm{R,inter-body}}(t)$ denotes the corresponding R-peak amplitude in the inter-body ECG signal. To mitigate fluctuations caused by beat-to-beat variability and interference from the other participant’s cardiac activity, R-amplitude ratios were averaged over 10 seconds. Ratios were computed for each participant regardless of whether they performed the mouth contact or served as the receiving body.

\begin{figure*}[h!]
  \includegraphics[width=\textwidth]{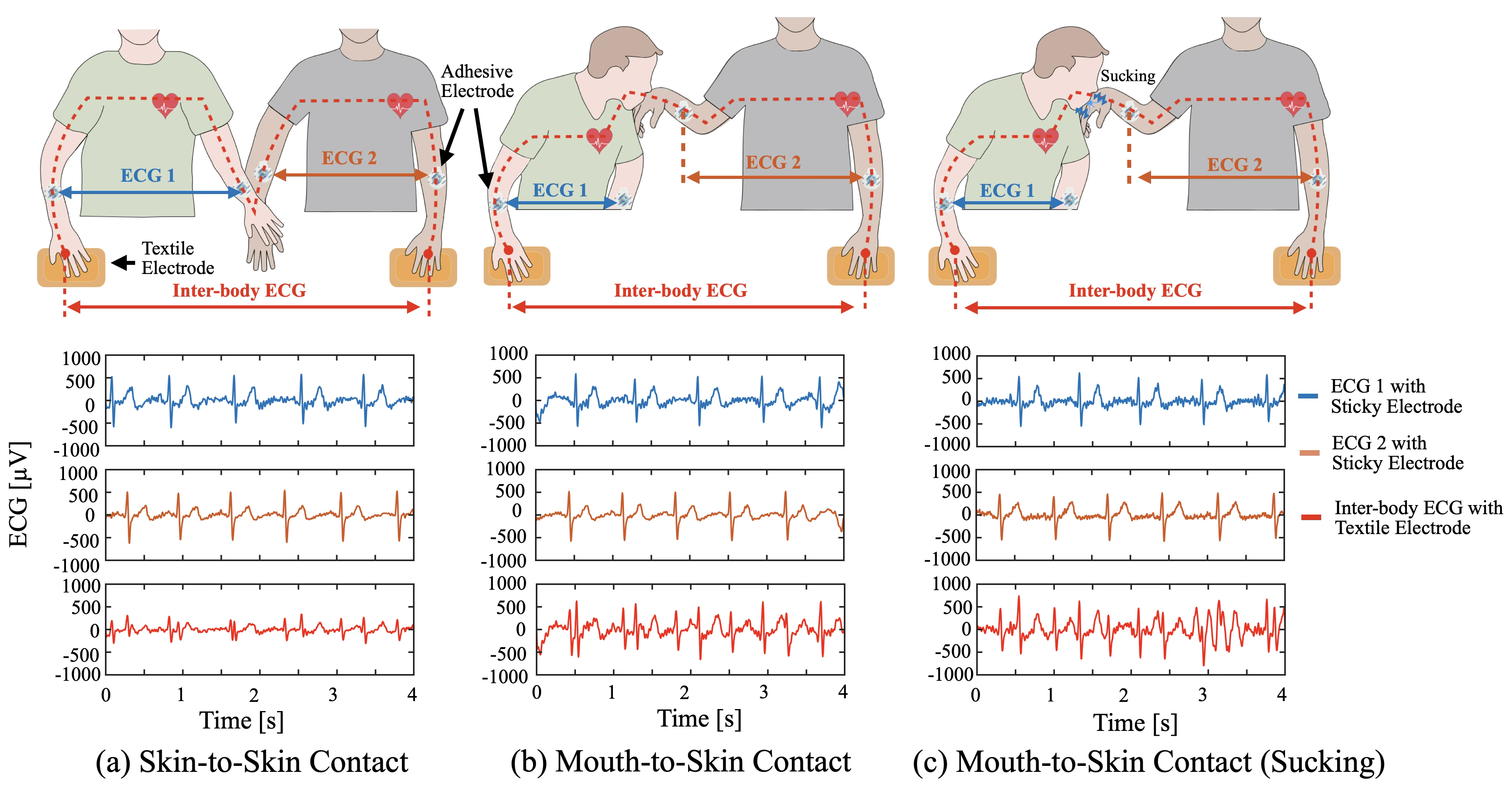}
  \caption{Study setup and exemplary ECG signals recorded in Experiment 1 under three contact conditions: (a) general skin-to-skin contact (hand holding), (b) static mouth-to-skin contact, and (c) mouth-to-skin contact with rhythmic sucking motions.}
  \Description{}
  \label{fig:exp1}
\end{figure*}
In addition to that, we also evaluated whether inter-body ECG signals remain stable during movement (e.g., rhythmic sucking). To quantify this, we assessed R-peak detection accuracy using Pan-Tompkins algorithm~\cite{pan} under the sucking condition. Note that reliable R-peak detection is important because R peaks serve as the primary reference for deriving key physiological metrics, including heart and respiration rate. Respiration is inferred through respiratory sinus arrhythmia and the modulation of R--R intervals~\cite{bental2012evaluating}.
Thus, stable R-peak detection under the sucking condition can indicate that inter-body ECG signals maintain sufficient signal quality for downstream physiological inference.

\textbf{Results.} Our results show that individual ECG signals can be transmitted through all three tested contact conditions. Notably, inter-body ECG coupling is significantly stronger during mouth-to-skin contact (R-amplitude ratio: mean = 0.91, SD = 0.03) compared to general skin-to-skin contact (mean = 0.64, SD = 0.07). \rev{In addition, the mouth-to-skin R-amplitude ratio remained similar with and without sucking (with sucking: 0.94; without sucking: 0.91). This suggests that oral physical contact itself, rather than active sucking, is the primary factor contributing to stronger coupling. Therefore, the R-amplitude ratio may serve as a useful indicator for identifying latch onset during breastfeeding.}

For R-peak detection, we found that rhythmic sucking introduces observable temporal fluctuations in signal amplitude but does not substantially degrade detection performance (F1 score mean= 0.94, SD = 0.06). Fig.~\ref{fig:exp1} illustrates an example from one participant dyad. During general skin-to-skin contact (Fig.~\ref{fig:exp1}a), the mixed inter-body ECG exhibits low-amplitude R peaks, indicating weaker electrical coupling. In contrast, during mouth-to-skin contact (Fig.~\ref{fig:exp1}b), the transmitted R peaks are noticeably larger and more distinct, reflecting stronger coupling and lower contact impedance. Even when rhythmic sucking is present, causing visible fluctuations in baseline and peak amplitude, the majority of R peaks remain detectable, supporting the inter-body ECG sensing under realistic feeding-related motion.

\subsection{Experimental Study 2 - Inter-body Acoustic Transmission}\label{exp2_frquency_range}

The objective of this study is to understand how acoustic signals, generated by sucking and swallowing, propagate across the inter-body contact interface. Prior research has demonstrated that these events can be reliably detected using contact microphones placed on the user's cheek \cite{BodyBeat, 10.1007/11551201_4} and throat \cite{NeckSense}. In our work, we aimed to compare the acoustic signals captured after transmission through the mouth-to-breast contact with those recorded directly from a microphone attached to the throat, which served as the ground-truth reference. This comparison allows us to evaluate how much of the original acoustic signature remains after transmission through soft tissue and inter-body coupling.

\textbf{Experiment Setup.} Similar to the ECG study, we did not conduct this experiment with real caregiver--infant dyads \rev{due to ethical and practical constraints. Instead, we used a silicone artificial breast (Vollence, Silicone Breast~\cite{vollence_breastforms_amazon_2026}) as a controlled and repeatable benchtop testbed for preliminary characterization. This homogeneous model cannot fully capture the acoustic damping behavior of human breast tissue. However, existing breast phantoms are primarily designed for ultrasound imaging rather than audible-band body-sound propagation during breastfeeding. Given this limitation, the silicone model is a reasonable alternative because it provides a soft, deformable, and lossy medium with a breast-like form factor. Similar silicone models are commonly used as educational and training phantoms~\cite{USTBAS20189,sadovnikova2020development,GALLAWAY2025e00707}.}

\begin{figure*}[h!]
  \includegraphics[width=\textwidth]{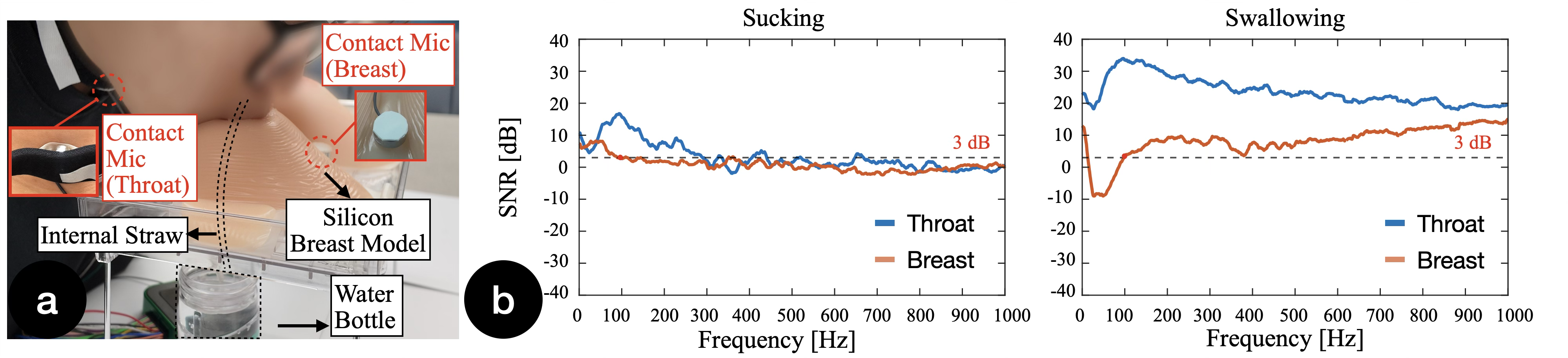}
  \caption{(a) Experiment 2 setup. (b) Average SNR of sucking and swallowing sounds via inter-body transmission.}
  \Description{}
  \label{fig:exp2}
\end{figure*}

We recruited 5 adult participants (3 males, 2 females, aged 21-27) and instructed each participant to perform three consecutive actions 10 times on the model: (1) hold, (2) suck, and (3) swallow. Because direct swallowing on the silicone breast does not involve fluid intake and thus may not fully reproduce natural swallowing acoustics, we modified the model by integrating a water bottle (Fig.~\ref{fig:exp2}a) that allows participants to actually drink water during the task.
To collect acoustic signals, we attached one contact microphone (TE Connectivity, CM\,-01B~\cite{CM01B}) to the participant's throat to capture ground-truth swallowing acoustics, and placed a second contact microphone on the silicone breast to record the cross-body transmitted acoustic signals. Both microphones were connected to an oscilloscope (Analog Discovery 3 \cite{digilent_analog_discovery_3}) sampling at 2000 Hz.  This setup allows direct comparison between the original swallowing and sucking sounds and the attenuated signals measured through the inter-body interface.

\textbf{Data Analysis.} All acoustic signals were first high-pass filtered \rev{(a 4th-order Butterworth high-pass filter with cut-off frequency of 15~Hz)} to remove DC offset and low-frequency motion artifacts. We then manually segmented the signals into individual event windows corresponding to holding, sucking, and swallowing actions, using a fixed window size of 500~ms centered around each event onset. For each segmented window, we computed the frequency-domain representation using a Fast Fourier Transform (FFT) to obtain the power spectrum. Then, we then quantified signal-to-noise ratio (SNR) by comparing the spectral power of sucking or swallowing events to that of holding events, which served as a baseline noise reference. SNR values were then averaged across participants to characterize the acoustic signatures of sucking and swallowing.

\textbf{Results.} Fig. \ref{fig:exp2}b shows the average signal-to-noise ratio (SNR) for sucking and swallowing across the 0–1000 Hz band, computed from the two microphone placements (breast-mounted vs. throat-mounted). Despite attenuation relative to throat measurements ($\Delta$SNR = \rev{-8.30} dB for sucking and $\Delta$SNR = -11.64 dB for swallowing), both sucking and swallowing signals remain observable at the breast model. Sucking exhibits a predominantly low-frequency signature (< 100 Hz). However, its spectral characteristics, especially between 0--60 Hz, slightly differ between breast and throat measurements because sucking directly applies mechanical pressure to the breast, inducing local tissue vibration in addition to bio-acoustic signals originating from the infant’s mouth. In contrast, swallowing shows higher-frequency components (100–1000 Hz), with more similar spectral profiles between throat and breast measurements \rev{($r = 0.84$)}, suggesting that swallowing-related acoustic vibrations propagate through the oral structures, nipple, and breast model. These results indicate the feasibility of capturing sucking and swallowing related events from breast-mounted acoustic sensing, supporting inter-body acoustic transmission as a viable sensing mechanism.

\subsection{Experiment Summary and Sensing Challenges}
Through the two experiments, we verify three key assumptions. First, mouth-to-skin contact indeed results in stronger electrical coupling than general skin-to-skin contact, with higher R-peak amplitude ratios between individual and inter-body ECGs. Second, the rhythmic movements induced by sucking though introduce temporal fluctuations in inter-body ECG amplitude, but do not disrupt R-peak detectability, indicating that core cardiac features remain observable under realistic feeding-related motion. Third, acoustic signals generated by sucking and swallowing remain observable on the caregiver’s body despite attenuation.

These findings also reveal three sensing challenges that must be addressed for the caregiver-worn sensing system.  
\begin{itemize}
    \item Although R-peak amplitude ratios can be used to detect mouth-to-skin contact, reliably establishing an inter-body electrical pathway during natural breastfeeding remains challenging. This requires ergonomically designed garments that facilitate consistent contact between the infant’s body (e.g., hand or foot) and at least one ECG electrode.
    \item While sucking-related motion does not disrupt R-peak detectability, separating infant ECG from inter-body ECG remains challenging due to baseline and amplitude fluctuations induced by rhythmic sucking.
    \item Because sucking and swallowing signals are attenuated at the caregiver’s body, reliably identifying their acoustic signatures under noise and motion is also challenging.
\end{itemize}


\section{MAMMAL DESIGN}

To address the challenges, we design Mammal, a hardware–software co-design solution for noninvasive breastfeeding monitoring. In the following sections, we first describe the hardware design of the garment and then explain how the system achieves the sensing objectives through analysis of the collected signals.

\subsection{Garment Design}

The design of the Mammal garment is guided by three primary design goals: 
(1) to avoid placing sensors or circuit boards directly on the infant by embedding all sensing components into a caregiver-worn garment; 
(2) to ensure sensing and ergonomic comfort across diverse breastfeeding postures and contexts; and 
(3) to preserve a familiar nursing-garment form factor that supports natural, unobstructed breastfeeding routines.

\begin{figure}
  \includegraphics[width=\textwidth]{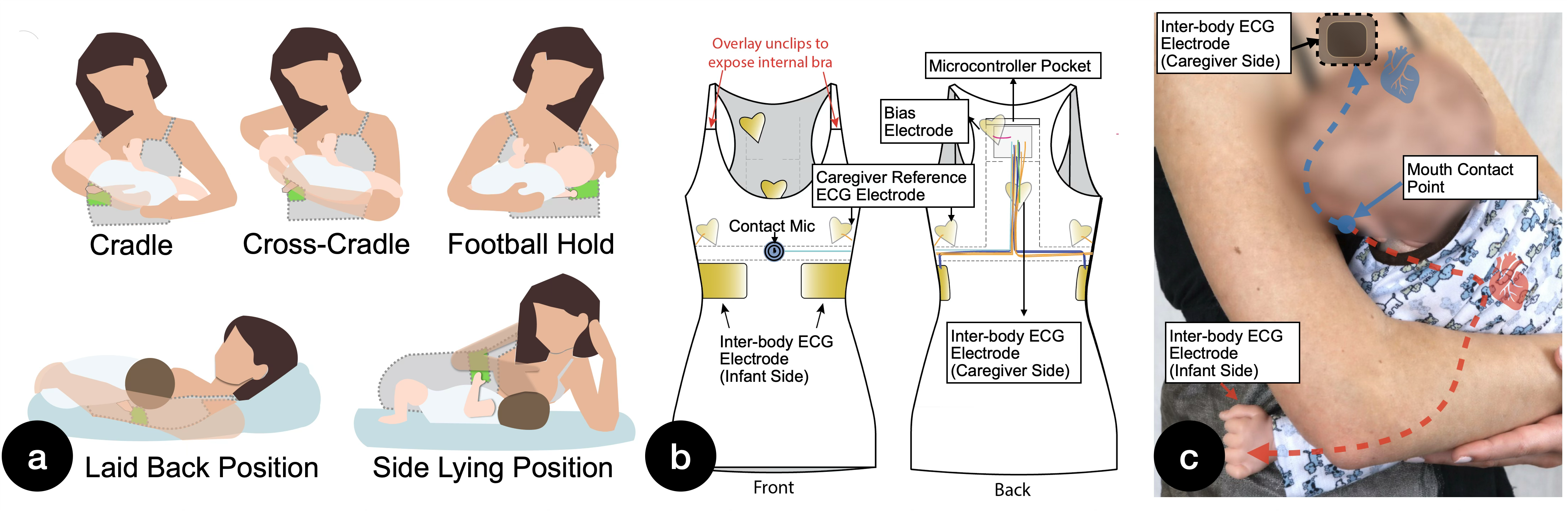}
  \caption{Garment design and sensor placement.
(a) Five common breastfeeding postures, highlighting infant contact near the waist (green).
(b) Mammal garment design and sensor layout (front/back).
(c) Example of natural infant hand contact near the waist during breastfeeding \rev{and mouth--skin contact. Blue and red dashed lines schematically indicate the caregiver-side and infant-side inter-body ECG electrical pathways, respectively.}}

  \Description{}
  \label{fig:garment_design}
\end{figure}

To meet these design goals, we first examine how breastfeeding is typically performed. We conducted interviews with three breastfeeding mothers and reviewed established lactation guidelines \cite{westerfield2018breastfeeding}, deriving three design considerations:
\begin{itemize}
    \item  \textbf{Supporting Five Feeding Postures.} 
    There are five common breastfeeding postures, including cradle, cross-cradle, football, laid-back, and side-lying. They are illustrated in Fig.~\ref{fig:garment_design}a. Caregivers usually choose one posture at the beginning of a feeding session based on personal preference and situational needs (e.g., available cushions or seating). Once selected, the posture is usually maintained for the entire session to keep the infant settled and avoid disrupting the latch. Accordingly, our garment must support sensing across all five postures without requiring repositioning of sensors during feeding.

    \item \textbf{Accommodating Left–Right Breastfeeding.}  
    Caregivers commonly alternate between the left and right breast across feeding sessions to balance milk supply and reduce nipple discomfort. During each session, only the nursing breast is typically exposed, while the rest of the torso remains covered for comfort, warmth, and privacy. Therefore, the garment must support sensing on either breast without requiring manual repositioning or additional exposure of the caregiver’s body.

    \item \textbf{Working with Limited Infant Skin Exposure.} Depending on temperature and context, many infants breastfeed while wearing clothing, especially out of home, leaving only the head, hands, and feet uncovered. As a result, the exposed electrode for inter-body ECG sensing must be carefully positioned to ensure natural contact with the expose infant skin.
\end{itemize}

Based on these considerations, we design a garment prototype in the form factor of a nursing tank top, as shown in Fig.~\ref{fig:garment_design}b. This garment prototype integrates three sensing channels:

\textbf{Inter-body ECG channel.} This channel uses a pair of textile electrodes to capture inter-body ECG signals. One electrode is exposed to enable natural contact with the infant’s skin, while the other rests against the caregiver’s torso to form a conductive pathway across both bodies. To promote infant contact regardless of breastfeeding side, we place exposed electrodes symmetrically on the left and right waist. This placement is informed by our posture analysis, which shows that across common breastfeeding postures the infant’s hands frequently rest or move in these regions while grasping the caregiver’s clothing. Because the hands are often uncovered, these interactions create opportunities for skin contact (Fig.~\ref{fig:garment_design}c). The other measurement electrode is positioned on the caregiver’s upper back, adjacent to the micro-controller so the electrical pathway traverses both bodies and enables inter-body ECG capture during feeding.
\rev{ In addition, a bias electrode is placed near the caregiver's shoulder blade and connected to the Cyton BIAS output for common-mode noise rejection. Unlike the measurement electrodes, this electrode acts as a driven-bias feedback electrode to helps reduce common-mode noise and stabilize body potential during acquisition.}

\textbf{Caregiver-reference ECG channel.}   
Because the inter-body ECG channel contains both infant ECG and caregiver ECG, we include a caregiver-reference ECG channel to measure the caregiver-only component for ECG separation and noise canceling.
This channel uses another pair of textile electrodes placed above the left and right waist near the lateral breast. These locations are above the exposed inter-body ECG electrodes, yielding similar conductive-trace routing. As a result, conductive-line noise is captured in the caregiver-reference ECG measurements, supporting suppression of shared noise components when separating infant ECG from inter-body signals.

\textbf{Acoustic sensing channel.}
To capture inter-body acoustic signals, a small rigid contact microphone is placed at the lower central region of the breast to support both left- and right-side breastfeeding. This region often aligns with structured or lightly padded areas of nursing bras, reducing wearer discomfort. To improve acoustic coupling, the microphone is integrated into a soft, compressive elastic band~\cite{band} that maintains gentle, continuous contact with the skin. The band can be routed through the bra’s lower cup seam to provide stable pre-load without pressure points, enabling consistent capture of sucking and swallowing vibrations.

All sensing channels are routed along the underbust seam to the upper back and connect to a microcontroller housed in an upper-back pocket, keeping the front unobstructed for caregiver–infant contact during breastfeeding.

\subsection{ECG Separation and Noise Canceling}

\begin{figure}
  \includegraphics[width=\textwidth]{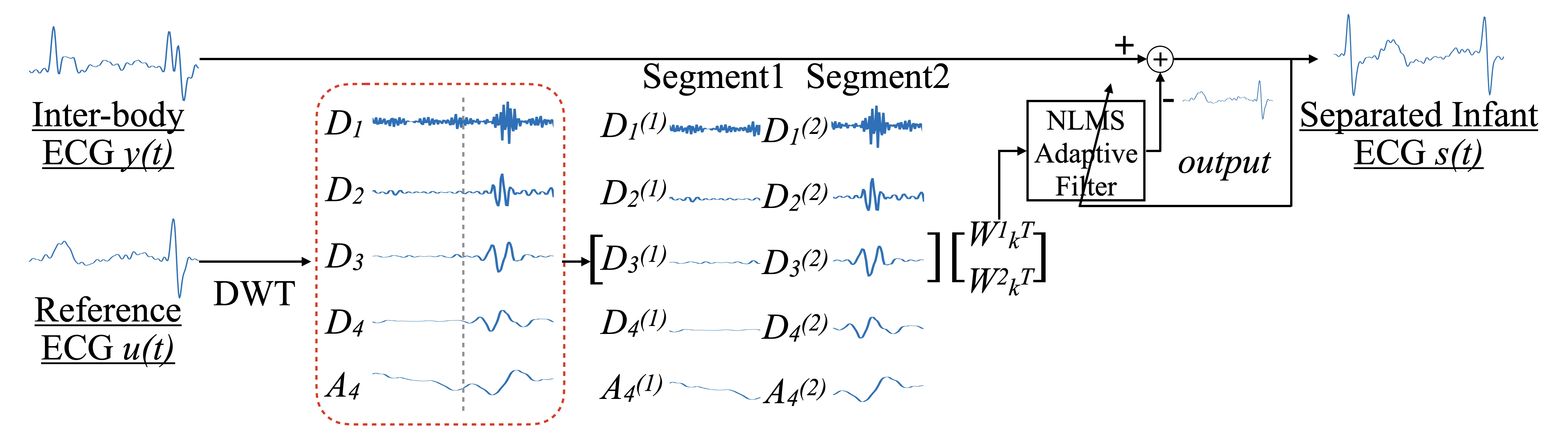}
  \caption{Overview of the ECG separation pipeline.}
  \Description{}
  \label{fig:ECG_separation}
\end{figure}

With the garment design, our next step is to develop an algorithm to separate the infant ECG component from the inter-body ECG signal (Fig.~\ref{fig:ECG_separation}). Following the signal model introduced in prior work~\cite{shao2024joey}, we express the infant ECG component (\( s_{\text{infant}}(t) \)) as:
\begin{equation}
s_{\text{infant}}(t)= y_{\text{mix}}(t) - w * u_{\text{caregiver}}(t) - \eta_{\text{line}}(t) - \eta_{\text{contact}}(t),
\end{equation}
where \( y_{\text{mix}}(t) \) denotes the inter-body mixed ECG signal measured on the garment, \( u_{\text{caregiver}}(t) \) represents the theoretical caregiver ECG component contributing to the \rev{inter-body} signal (rather than a directly captured reference ECG), \( w \) models the coupling strength of caregiver ECG into the \rev{inter-body} signal, \( \eta_{\text{line}}(t) \) is noise introduced by conductive traces and wiring in the garment, and \( \eta_{\text{contact}}(t) \) represents other motion-induced contact noise during caregiver–infant interaction.

In practice, the caregiver ECG measured on the garment is not a clean reference signal (\( u_{\text{caregiver}}(t) \)), but contains conductive-line noise profile (\( \eta_{\text{line}}(t) \)). In addition, different ECG components at distinct frequency bands experience different inter-body coupling strengths across the caregiver–infant contact, making a single scalar weight \( w \) not enough to model the process. To address this challenge, a typical approach is to decompose the measured caregiver ECG signal into multiple frequency subbands using a discrete wavelet transform (DWT), and to assign independent weights to each subband to model their respective contributions to the inter-body ECG signal. 
Accordingly, the infant ECG can be represented as:
\begin{equation}
s_{\text{infant}}(t)= y_{\text{mix}}(t)
- 
\big[ D_1(t), D_2(t), D_3(t), D_4(t), A_4(t) \big] \mathbf{w}^{\top}
- \eta_{\text{contact}}(t),
\end{equation}
where \( \{D_1, D_2, D_3, D_4, A_4\} \) denote the detail and approximation component obtained from a multi-level DWT of the caregiver ECG component measured on the garment, and \( \mathbf{w} \) is a vector of subband-specific weights. 

However, during breastfeeding, rhythmic sucking and body movement further induce time-varying changes in inter-body coupling. To account for this nonstationarity, we divide each 1-second analysis window into \( n \) shorter temporal segments and estimate time-local subband weights for each segment:
\begin{equation}
s_{\text{infant}}(t)= y_{\text{mix}}(t)
- \sum_{k=1}^{n} 
\big[ D_1^{(k)}(t), D_2^{(k)}(t), D_3^{(k)}(t), D_4^{(k)}(t), A_4^{(k)}(t) \big] \mathbf{w}_k^{\top}
- \eta_{\text{contact}}(t),
\end{equation}
where \( \mathbf{w}_k \) denotes the subband weight vector corresponding to the \(k\)-th temporal segment and \(n\) is 2 in our implementation based on empirical results.
To track time-varying inter-body ECG coupling dynamics during feeding, these weights are updated online using an adaptive filter based on the Normalized Least Mean Squares (NLMS) algorithm \cite{205720}. \rev{This update allows the system to enable a more feeding-robust estimate of the infant ECG.}

\rev{\textbf{DeScoD-ECG Denoising.} Despite temporal-segmented adaptive separation, the estimated infant ECG may still contain residual motion-induced contact noise \( \eta_{\text{contact}}(t) \). To suppress this noise, we apply a deep learning--based denoising model, DeScoD-ECG \rev{\cite{10018543}}. Unlike a fixed filter on the separated waveform, DeScoD-ECG learns the conditional distribution of clean ECG signals from noisy observations and reconstructs cleaner signals through an iterative reverse diffusion process. 
To avoid unnecessary denoising, we selectively activate DeScoD-ECG only when the short-time energy of the separated infant ECG, computed over a 1~s window, exceeds the average energy of the caregiver ECG within the same window. Empirically, under normal coupling conditions, the infant ECG component observed in the \rev{inter-body} signal is typically weaker than the caregiver ECG. Therefore, if the separated infant ECG has unusually high energy relative to the caregiver reference, it is likely to contain substantial residual contact-related contamination rather than purely infant cardiac activity. When this condition is met, we apply the adapted DeScoD-ECG model to 0.5~s windows to suppress morphology-corrupting residual noise while preserving local heartbeat structure. }

\subsection{Sucking and Swallowing Detection}
\begin{figure}
  \includegraphics[width=\textwidth]{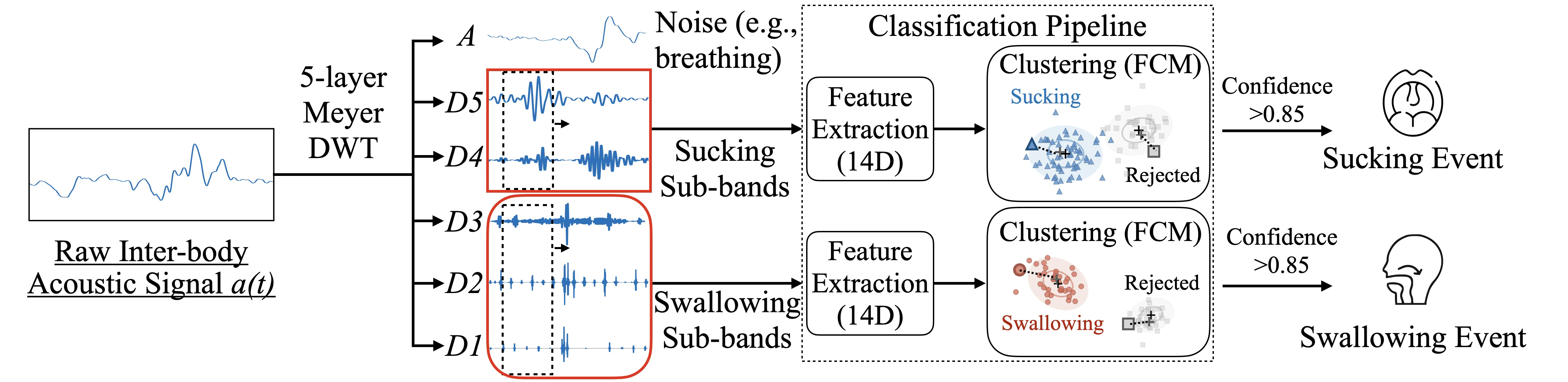}
  \caption{Overview of the algorithm for sucking and swallowing detection.}
  \Description{}
  \label{fig:alg}
\end{figure}

To estimate the SSB ratio, one of the key sensing objectives, we must not only infer infant ECG but also identify sucking and swallowing events from inter-body acoustic signals. Based on findings from Experiment~2, which reveal distinct and separable frequency ranges for sucking and swallowing acoustics, we design a multi-stage signal processing pipeline that leverages frequency-domain separation and event-level analysis, as shown in Fig.~\ref{fig:alg}.

In this pipeline, we first apply a level-5 Meyer Discrete Wavelet Transform to decompose the inter-body acoustic signal \(a(t)\) into multiple frequency subbands, enabling separation of physiologically distinct acoustic components based on their frequency spectral characteristics:
\begin{equation}
a(t) = \sum_{j=1}^{5} D_j(t) + A_5(t),
\end{equation}
where \(D_j(t)\) denotes the detail coefficients at level \(j\) and \(A_5(t)\) represents the lowest-frequency approximation component.


The lowest-frequency approximation subband \(A_5\), capturing frequencies below approximately 16~Hz, is excluded from further analysis, as it is dominated by low-frequency physiological activity such as heart sounds and respiration-induced body motion, which are not informative for feeding event detection.

Among the remaining detail subbands, \(D_4\) and \(D_5\) primarily capture mid-frequency components, corresponding approximately to the range 31–125~Hz, which are associated with rhythmic oral–motor movements during feeding. As shown in Experiment~2, these subbands exhibit frequency signatures that align with infant sucking behavior and are therefore used for sucking event detection. In contrast, higher-frequency subbands \(D_1\), \(D_2\), and \(D_3\), spanning frequencies above approximately 125~Hz, emphasize short-duration, transient acoustic events with sharper temporal signatures. These characteristics are consistent with swallowing sounds produced by rapid bolus transfer and airway closure, and these subbands are thus used for swallowing event detection.

To enable robust event detection, we adopt a feature-driven online fuzzy clustering pipeline ~\cite{dudik2015comparative}. \rev{This pipeline avoids fixed thresholds, which are brittle when signal amplitude varies with sucking strength and caregiver movement. 
Specifically, we analyze sucking and swallowing separately within their respective frequency bands. Within each band, we slide a 0.3-s rolling window over the reconstructed subband signal and compute a 14-dimensional feature vector for every window. The feature set includes time-domain statistics (e.g., peak amplitude, duration, temporal symmetry), spectral descriptors (e.g., subband energy, spectral centroid), and wavelet-based measures (Table~\ref{tab:features}).  These features provide a more stable representation than raw amplitude alone.
To translate features into detections, we maintain two fuzzy clusters,} \rev{respectively for sucking and swallowing detection}.  
The cluster centers are initialized using 20 manually annotated sucking and swallowing samples collected from a pilot caregiver–infant dyad.  As each window arrives, we compute its soft membership to both clusters, yielding a continuous event-confidence score. We classify the window as an event when its maximum membership exceeds 0.85. \rev{This value was selected empirically in pilot study and then held constant across all sessions and participants.}
\rev{Otherwise, it is treated as motion-induced noise. After scoring each window, we update the cluster centers incrementally using the window’s memberships. This online adaptation allows the model to track gradual changes in artifact level and signal strength without session-specific labeling, manual recalibration, or threshold adjustment. As a result, the detection pipeline generalizes across participants and sessions, while still automatically adapting to the signal characteristics of the ongoing session.}


\begin{table}[h]
\centering
\caption{Summary of the 14-dimensional feature vector. Features are grouped by domain to capture temporal, spectral, and morphological characteristics.}
\label{tab:features}

\renewcommand{\arraystretch}{0.9} 
\small 

\resizebox{\columnwidth}{!}{%
\begin{tabular}{@{} l p{4.5cm} p{7cm} @{}}
\toprule
\textbf{Domain} & \textbf{Features} & \textbf{Rationale} \\ \midrule

\textbf{Time} 
& RMS, Crest Factor, ZCR, Duration, Skewness, Kurtosis 
& Captures signal energy, impulsiveness, noisiness, and probability distribution shape. \\ \midrule

\textbf{Frequency} 
& Spectral Centroid ($F_c$), Peak Freq, Bandwidth 
& Distinguishes low-freq bolus sounds from friction; identifies dominant spectral components. \\ \midrule

\textbf{Time-Freq} 
& Wavelet Energy Ratios, Wavelet Entropy ($WE$) 
& Measures energy distribution across sub-bands and signal complexity. \\ \midrule
\textbf{Periodicity} 
& ACF Peak Magnitude, Burst Regularity (CV) 
& Quantifies rhythmicity and rejects periodic noise (e.g., mains hum). \\ \bottomrule

\end{tabular}%
}
\end{table}

\subsection{Physiological and Feeding State Inference}

\textbf{Latching Onset and Duration.}
To identify latch onset and estimate latch duration, we compute the average R-amplitude ratio between the caregiver ECG and the inter-body ECG. When this ratio exceeds an empirically determined threshold (0.8), we infer the presence of mouth-to-breast contact and generate a binary latch-state signal indicating latch (1) or no latch (0). The transition from 0 to 1 is marked as the latch onset.
To estimate latch duration, we apply a temporal morphological closing operation with a structuring element of 0.8~s (determined from pilot study~\cite{shao2024joey}) to the binary latch-state signal. This operation suppresses spurious state transitions caused by brief signal dropouts or motion-induced noise while preserving sustained latch periods. The final latch duration is computed as the total length of contiguous segments in the resulting smoothed latch-state signal.

\textbf{Heart Rate Inference.} 
We compute the infant’s heart rate from the separated infant ECG signal. Specifically, we apply the well-established Pan--Tompkins algorithm to detect QRS complexes in the ECG signal~\cite{pan, heartrategarment}. The detected R-peaks are then used to compute inter-beat (RR) intervals. The instantaneous heart rate is calculated using a sliding window of ten consecutive RR intervals as:
\begin{equation}
H = \frac{60}{\frac{1}{10}\sum_{i=1}^{10} d_i},
\end{equation}
where \( d_i \) denotes the duration (in seconds) of the \(i\)-th RR interval, and ten RR intervals are used to detect heart rate changes in infants.

\del{In addition to average heart rate, we also compute heart rate variability (HRV) metric from the RR interval sequence to characterize beat-to-beat fluctuations. Specifically, we calculate the standard deviation of RR intervals (SDNN) as:
$\text{SDNN} = \sqrt{\frac{1}{N-1}\sum_{i=1}^{N} \left(d_i - \bar{d}\right)^2},$
where \( d_i \) denotes the \(i\)-th RR interval, \( \bar{d} \) is the mean RR interval within the analysis window, and \( N \) is the number of RR intervals used for computation.}

\rev{In addition to average heart rate, we also compute the heart rate variability (HRV) metric from the RR interval sequence to characterize beat-to-beat fluctuations. Specifically, we calculate the root mean square of successive differences (RMSSD) as:
\begin{equation}
\mathrm{RMSSD} = \sqrt{\frac{1}{N-1}\sum_{i=1}^{N-1}\left(d_{i+1}-d_i\right)^2},
\end{equation}
where \( d_i \) denotes the \(i\)-th RR interval, and \( N \) is the number of RR intervals used for computation.}

\textbf{SSB Ratio Estimation.}
The SSB ratio is a commonly used indicator of an infant’s neuromuscular coordination during breastfeeding. In Mammal, we estimate SSB by combining (1) sucking and swallowing events from inter-body acoustics and (2) breathing events inferred from respiration-induced modulation of the separated infant ECG.
To estimate breathing from the separated infant ECG, we exploit the fact that respiration modulates ECG through mechanisms such as baseline wander, amplitude modulation, and frequency modulation (FM). We adopt an FM-based approach that leverages respiration-linked modulation of cardiac timing. We detect R-peaks from the separated ECG, compute the inter-beat-interval (IBI) series $\mathrm{IBI}_i = t_{i+1}-t_i$, resample the IBI series to a uniform grid, and compute its spectrum (FFT). The dominant peak within a physiologically plausible infant respiration band (0.5--1.2~Hz) is selected as the breathing frequency $f_{\mathrm{resp}}$. We then apply \rev{a 4th-order Butterworth band-pass filter with cut-off frequencies of 0.5--1.2~Hz to the resampled IBI series. We identify breath events from the filtered IBI-derived respiratory waveform using local-maximum peak detection. The minimum peak-to-peak distance is set to \(1/1.2\)~s, corresponding to the upper bound of the infant respiration band (0.5--1.2~Hz), to avoid double-counting within one respiratory cycle. Each retained local maximum is treated as one breathing event.}

Given detected sucking, swallowing, and breathing events, we compute the instantaneous SSB ratio using breath-centered temporal windows. 
Specifically, for each detected breathing event at time $t_b$, we count the number of sucking and swallowing events occurring in the preceding 10\,s interval $[t_b-10,\,t_b)$. This yields a per-breath SSB triplet:
\begin{equation}
    \mathrm{SSB}(t_b) = N_{\mathrm{suck}}(t_b) : N_{\mathrm{swallow}}(t_b) : 1,
\end{equation}
where $N_{\mathrm{suck}}(t_b)$ and $N_{\mathrm{swallow}}(t_b)$ are the counts of sucking and swallowing events in $[t_b-10,\,t_b)$, and the breathing count is fixed to 1 for that breath-centered window. 

However, because breathing detection is based on rhythmic patterns, the instantaneous per-breath SSB ratio can be noisy. We therefore compute an aggregated SSB ratio by pooling event counts across multiple breaths within a longer interval (e.g., a 10~s window). We sum the detected sucking and swallowing events across all breath-centered windows in the interval and normalize by the total number of breaths, reporting the resulting average sucks-per-breath and swallows-per-breath as the SSB ratio.

\textbf{Breastmilk Intake Estimation.}
We estimate breastmilk intake by summing the milk transferred, using the established relationship between milk transfer and inter-suck interval (ISI) \cite{bowenjones1982milkflow}. \rev{Specifically, the ISI, defined as the time between consecutive sucks, is related to milk flow. Shorter ISIs are associated with little or no milk transfer (i.e. non-nutritive sucking), whereas longer ISIs tend to indicate greater milk transfer (i.e. nutritive sucking). Based on this relationship, we estimate milk transfer for each suck from its ISI and sum these estimates over the feeding session to obtain total breastmilk intake.}

Following the linear model in prior work \cite{bowenjones1982milkflow},
milk transferred per suck ($m$, g/suck) is related to ISI (ms) as:
\begin{equation}
    m = \frac{\min(\mathrm{ISI}, cap)-a}{b},
    \label{eq:milk_regression_inverted}
\end{equation}
where the prior work reported $a=578$, $b=776$, and the ISI $cap=1300$\,ms. We compute ISI over successive sucks within a session and apply the model to estimate milk transferred per suck.
Session-level milk intake is then estimated by aggregating per-suck transfer:
\begin{equation}
       M~(\mathrm{g}) = \sum_{i=1}^{N_{\mathrm{suck}}} m_i,
\end{equation}
where $N_{\mathrm{suck}}$ is the number of detected sucks. This approach provides breastfeeding-compatible estimation from caregiver-worn sensing without requiring bottle feeding or test-weighing during use.

\section{MAMMAL IMPLEMENTATION}

\textbf{Nursing Garment.}
We collaborated with professional apparel designers to implement a custom nursing tank top that integrates sensing components while preserving comfort and everyday wearability. The garment features a detachable front panel that can be opened using a double-clip mechanism, allowing caregivers to expose the breast for latching with a single hand. This design enables feeding access without lifting or removing the entire garment, helping maintain stable positioning of the embedded sensors during breastfeeding.

Textile electrodes are fabricated using knit conductive textile material~\cite{adafruit1167} (electrode area: 76.2~mm~$\times$~76.2~mm; thickness: 0.5~mm), selected for its skin compatibility, flexibility, and wash durability. Acoustic sensing is performed using a miniature contact microphone~\cite{CM01B}, which is attached to an elastic band~\cite{band} using a fabric tape to enhance mechanical coupling to the body.

Sensor transmission lines are routed using parallel zig-zag stitching with silver conductive thread~\cite{adafruit} along the underband region of the garment, which runs underneath the cups and around the torso. This routing strategy leverages a mechanically stable region of the garment, improving signal stability while minimizing wearer discomfort. The transmission lines terminate at a low-profile 10-pin flexible printed circuit (FPC) connector~\cite{digikey-psr1636-10} positioned on the back of the garment, providing a reliable electrical interface to the microcontroller unit (MCU). To further enhance comfort around the electronics, we added a 5~mm-thick soft foam backing beneath the MCU enclosure to distribute pressure and prevent localized discomfort during wear.

\textbf{Micro-Controller Unit.}
We use an off-the-shelf OpenBCI Cyton board as the main controller, powered by a 3.7~V, 500~mAh LiPo battery, for synchronized ECG and contact-microphone sensing. The system consumes approximately 185~mW (3.7~V, 50~mA) during continuous operation and can last for 10~hours. In the default wireless mode, the Cyton samples and wirelessly transmits data at 250\,Hz via the USB dongle receiver \rev{~\cite{Cyton_USB}}. This rate is sufficient for ECG monitoring but is too low to capture short acoustic events (e.g., swallowing). \rev{Therefore, the wireless mode is only used for real-time data visualization. For recording both ECG and acoustic signals, we deploy the SD-card logging firmware~\cite{openbci-cyton-sdcard} on the Cyton and increase the sampling rate of acoustic signals to 2000\,Hz,} which preserves signal components up to 1\,kHz under the Nyquist limit, which requires the sampling frequency to be at least twice the target signal frequency.


We assign Cyton channels 1, 4, and 8 to the \rev{inter-body ECG} channel, the caregiver-reference ECG channel, and the contact microphone, respectively. The contact microphone is powered from the Cyton’s voltage output (Vout) and its analog output is sampled by the Cyton. \del{The garment’s ground electrode is connected to the Cyton ground to provide a stable reference for acquisition.}
\rev{The garment’s bias electrode (electrode area: 80mm x 80mm thickness: 0.5mm) was connected to the Cyton BIAS output to provide driven common-mode feedback and reduce common-mode noise during ECG acquisition.}

\textbf{Analysis.}
We implement a post-processing pipeline in MATLAB to synchronize and transform raw sensor signals into session-level metrics.

\section{EVALUATION}
The goal of this study is to \rev{understand} the performance of Mammal with real caregiver-infant dyads.  We also aim to understand user comfort, wearability, and usability.

\subsection{Participant}
We recruited 10 caregiver--infant dyads. Caregivers ranged in age from 26 to \rev{39} years, and reported bra sizes ranged from 32C to \rev{44DDD}. Infants ranged in age from 2 to 15 months at the time of the study. \rev{6} were male and \rev{4} were female. All infants were born at term and were actively breastfeeding, with no known feeding or cardiopulmonary disorders. All procedures were approved by the institutional review board, and informed consent was obtained from caregivers. Each caregiver received a \$50 gift card in compensation for their participation. Detailed participant demographics are summarized in Appendix~\ref{demographics}.

\subsection{Procedure}
Because breastfeeding is private, all sessions were conducted in a dedicated private clinical research space. Upon arrival, caregivers were introduced to the study space and procedure, given an opportunity to ask questions and choose to sign the consent form or not. Caregivers were then instructed to wear the Mammal garment and adjust the shoulder strap to ensure proper fit and comfortable skin contact between the garment electrodes and the caregiver’s body. Caregivers were also instructed on how to use a mobile application to provide ground-truth annotations of latch events and sucking activity during the study.
The study comprises three sessions totaling 25--55 minutes.

\textbf{Session 1: Breastfeeding Preparation (5 minutes).}
During this session, caregivers were given time to prepare for breastfeeding in their usual manner. Preparation activities included weighing pre-feeding infants, selecting and uncovering the breast to feed from, choosing a comfortable breastfeeding posture, positioning the infant, and holding or calming the infant through hand or body. A breastfeeding couch and a long chair were available to accommodate different feeding postures (Appendix~\ref{studyroom}). The Mammal garment continuously collected data during this period to assess potential false-positive latch detections prior to the onset of active breastfeeding.

\textbf{Breastfeeding session (10--40 minutes).}
During this session, caregiver-infant dyads engaged in active breastfeeding. Breastfeeding duration varied from 16 to 33 minutes depending on infant state and maternal milk availability. Caregivers were free to pause breastfeeding or switch breasts as needed. Throughout the session, caregivers used a dedicated mobile application to annotate latch onset and infant sucking events by pressing on-screen buttons. Caregivers could request assistance from experimenters outside the room at any time if needed.

\textbf{Interview session (10 minutes).}
Following the breastfeeding session, caregivers continued to wear the Mammal garment but covered the breast using the garment’s double-clip mechanism. They first measured post-feeding weight for infants and then participated in a semi-structured interview that assessed comfort, wearability, and usability, and gathered caregivers’ perceptions of the system during breastfeeding.

\subsection{Data Collection}
Throughout the study, the Mammal garment continuously collected raw signals from inter-body ECG channel, caregiver-reference ECG channel, and embedded acoustic sensor. To obtain ground-truth data, we used a multi-modal strategy that combined (1) caregiver manual annotation, (2) external monitoring, and (3) an infant-worn sensing device. Specifically, caregivers used the mobile application to record latch onset and duration, as well as infant sucking events, during breastfeeding. This is because caregivers were directly able to perceive latch and sucking through physical contact. Swallowing events were captured via external audio monitoring using a shotgun microphone (\rev{R{\O}DE NTG5~\cite{rode_ntg5}}) pointed to infant's head and neck. \rev{Because the room was equipped with noise control and designed to simulate a hospital-style clinical research environment, the microphone was able to capture swallowing sounds clearly with minimal environmental artifacts.} We adopted this approach because placing sensors around the infant’s head or neck was not permitted without additional IRB risk assessment, and video recording was not allowed due to privacy considerations. Swallowing events were identified using a state-of-the-art audio-based swallowing detection algorithm \rev{\cite{so2026swallowdetection, yamnet_tensorflow}}, and events with low confidence were subsequently reviewed and confirmed by a trained medical student. For infant ground-truth heart rate and breathing rhythm, we used an open-source Photoplethysmography (PPG)-based smartwatch (Bangle.js~\cite{banglejs-website}) worn on the infant’s ankle. We chose this device because it is more comfortable and infant-friendly than clinical ECG monitors, and because commonly used FDA-cleared infant monitors typically report only averaged values and do not provide fine-grained, time-domain respiratory signals needed to compute a ground-truth SSB ratio. By logging raw Bangle.js signals and analyzing them with respiratory-induced frequency variation (RIFV)~\cite{dehkordi2018extracting} algorithms, we can derive fine-grained pulse rate and breathing rhythm. For healthy individuals, these PPG-derived estimates closely track ECG-derived metrics~\cite{Watanabe}. To validate this setup, we conducted a calibration study with five adult participants wearing both the Bangle.js device and an FDA-approved pulse oximeter~\cite{masimo2019mightysat}. The results showed that heart rate and respiration rate measurements from the two devices differed by less than 1\%, consistent with prior work comparing Bangle.js with medical-grade sensors~\cite{Ravanelli2025OpenSourceSmartwatch}. Finally, for infant ground-truth intake, we used pre- and post-feeding infant weight measurements.

\subsection{Sensing Performance}
We examined the sensing performance of Mammal across all \rev{10} caregiver--infant dyads using the evaluation metrics described below. \rev{Table~\ref{tab:overall_performance_summary} summarizes Mammal's overall sensing performance across the 10 breastfeeding recordings. 
The table provides both ground-truth quantities and Mammal's estimated outputs for each sensing task, while detailed per-session results are reported in Appendix~\ref{appendix_per_session_results}.}

\begin{table*}[h!]
\centering
\caption{Overall sensing performance across the 10 breastfeeding recordings.}
\label{tab:overall_performance_summary}
\small
\setlength{\tabcolsep}{6pt}
\begin{tabular}{@{}p{3.4cm}p{5.0cm}p{6.2cm}@{}}
\toprule
\textbf{Sensing target} &
\textbf{Ground truth} &
\textbf{Overall performance} \\
\midrule

Latch onset detection &
$N_{\mathrm{GT}} = 102$ annotated latch onsets &
P/R/F1 = $0.93 \pm 0.07$ / $0.93 \pm 0.08$ / $0.93 \pm 0.05$ \\

Latch duration estimation &
Total GT latch duration = 239 min &
MAPE = $5.56 \pm 2.87\%$ \\

Infant heart-rate estimation &
PPG-derived infant HR trajectory &
MAE = $3.61 \pm 0.87$ bpm \\

Sucking detection &
$N_{\mathrm{GT}} = 9897$ annotated sucking events &
P/R/F1 = $0.91 \pm 0.03$ / $0.93 \pm 0.03$ / $0.92 \pm 0.03$ \\

Swallowing detection &
$N_{\mathrm{GT}} = 4283$ annotated swallowing events &
P/R/F1 = $0.92 \pm 0.03$ / $0.93 \pm 0.03$ / $0.93 \pm 0.03$ \\

\bottomrule
\end{tabular}
\end{table*}

\textbf{Latch onset and duration.}
Across \rev{10} breastfeeding sessions, we collected \rev{102} latch-onset events and \rev{239} minutes of breastfeeding data, along with \rev{101} minutes of skin-contact sessions and \rev{52} minutes of non-contact periods.

To account for temporal uncertainty introduced by human annotation latency, we applied a tolerance window when computing the event-level F1 score for latch onset detection. Specifically, latch onset detections were considered correct if they occurred within ±1.5 s of the caregiver-annotated timestamp. This tolerance corresponds to three consecutive detection windows (preceding, current, and subsequent), given the system’s 500 ms detection window.

\begin{figure*}[h]
    \centering
    \includegraphics[width=0.8\linewidth]{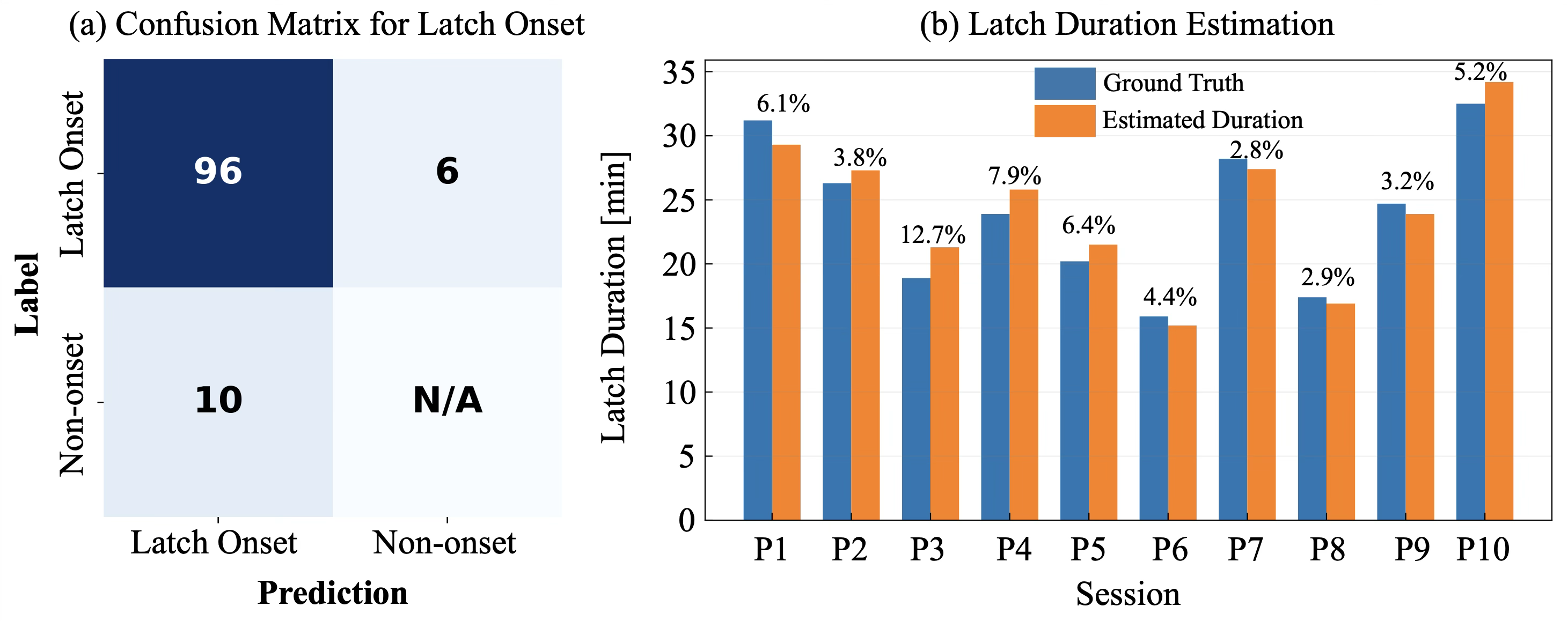}
\caption{Latch monitoring performance. 
(a) Confusion matrix for latch onset detection aggregated across the 10 sessions.
(b) Ground-truth and estimated latch durations for each session. 
The percentage above each pair indicates the absolute percentage error.}    \label{fig:latch_confusion_matrix}
\end{figure*}

Under this evaluation criterion, Mammal achieved a mean F1 score of \rev{0.93 (SD= 0.05)} for latch onset detection, with a precision of \rev{0.93 (SD= 0.07)} and a recall of \rev{0.93 (SD= 0.08)}. \rev{Fig.~\ref{fig:latch_confusion_matrix}a shows the confusion matrix of latch onset detection.}  During non-feeding periods, the system maintained a low false-positive rate of \rev{1.70\%}, indicating robust discrimination between latch and non-latch states. This is encouraging because multiple incidental skin contacts are common during breastfeeding preparation (e.g., caregivers holding the infant’s hands or the infant resting against the caregiver’s chest). The results suggest that mouth-to-breast contact produces a more distinct and stronger electrical coupling, even though concurrent skin-to-skin contacts may reduce contact impedance. 
Latch duration was estimated with a mean absolute percentage error of \rev{5.56\% (SD= 2.87\%)} across all feeding sessions. \rev{The duration error of each session is shown in Fig.~\ref{fig:latch_confusion_matrix}b.} Errors were generally stable, except in one session where the infant fell asleep mid-feed and intermittently broke and re-established latch. The caregiver noted that this pattern made manual annotation difficult, which likely contributed to the higher error in that session (\rev{12.70\%}).

\begin{figure*}[h]
    \centering
    \includegraphics[width=\textwidth]{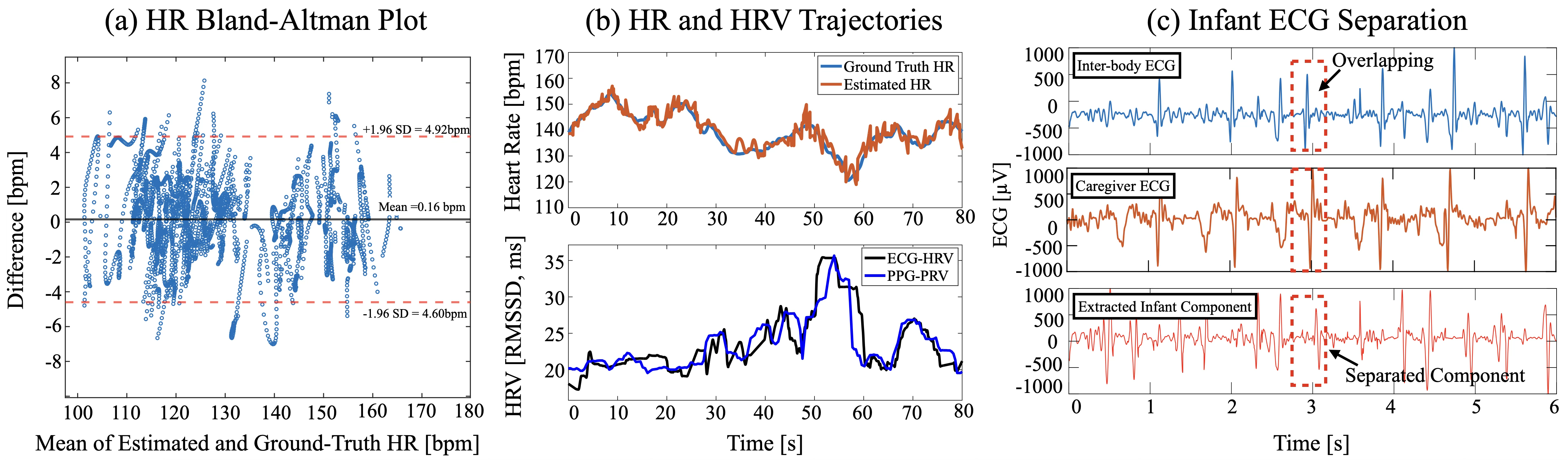}
  \caption{\rev{(a) Bland-Altman plot of Mammal-derived HR and the ground-truth HR from PPG.} (b) Mammal-derived \rev{HR and} HRV \del{(SDNN)} \rev{(RMSSD)} trajectories compared with PPG-derived \rev{heart rate and }variability. (c) Separated infant ECG shown alongside the \rev{inter-body} and caregiver ECG signals.}
    \label{fig:hrv}
\end{figure*}

\textbf{Infant heart rate estimation.}
Across all breastfeeding sessions, Mammal achieved a mean absolute error of \rev{3.61} bpm in infant heart rate estimation, in comparison with ground-truth measures derived from Bangle.js. This error falls within commonly used clinical tolerance bands for heart-rate monitoring (3-5~bpm) \cite{aami_ec13_2002}. 
To examine within-session performance beyond session-level averages, we further analyze time-varying heart rate and heart-rate variability (HRV) using sliding windows of 10 RR intervals. Fig.~\ref{fig:hrv} shows an example session comparing Mammal-derived heart rate (HR) and heart rate variability (HRV) with ground-truth, and illustrating the separated infant ECG alongside the inter-body ECG.  Note that the ground-truth trajectories are based on PPG, which instead reflects pulse rate and pulse-rate variability and may exhibit a small latency relative to Mammal. To quantify within-session agreement, we time-align the trajectories (allowing a constant lag per session) and then compute the Pearson correlation between Mammal and the PPG-based ground-truth trajectories over each session. As a result, Mammal achieved $r_{\mathrm{rate}}=\rev{0.95}$ and $r_{\mathrm{HRV}}=\rev{0.96}$, indicating that rate dynamics and variability trends remain consistent and aligned with ground-truth trajectories throughout feeding despite sucking-related motion and infant repositioning. These values are in line with prior reports showing strong agreement between ECG-derived HRV and PPG-derived PRV \cite{ajtay2023oscillating,Watanabe,mishra}.

\begin{figure}[h]
    \centering
    \includegraphics[width=\textwidth]{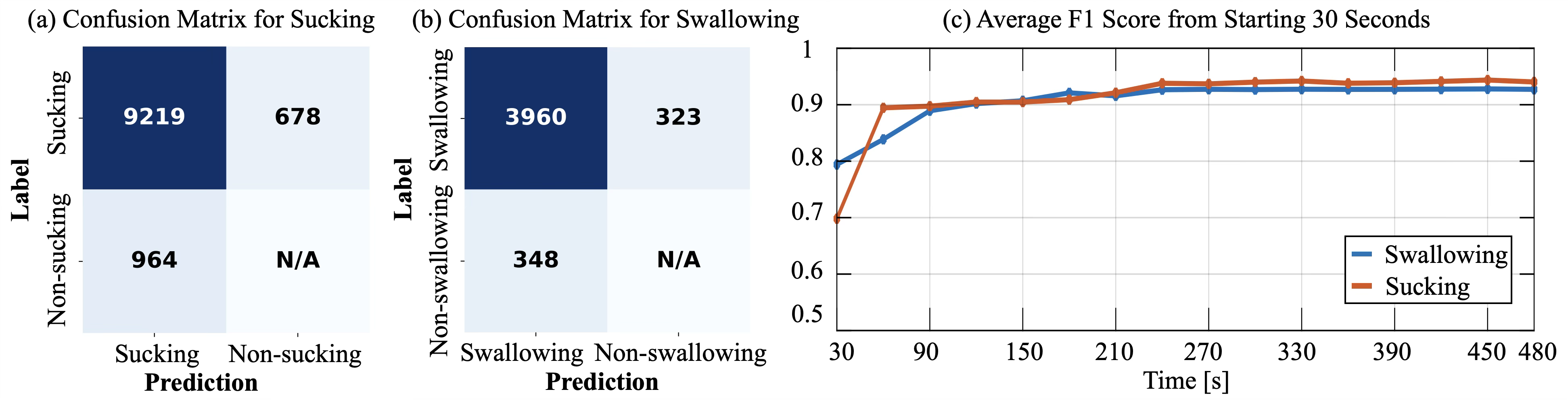}
    \caption{ (a) Confusion matrix for sucking detection. (b) Confusion matrix for swallowing detection. (c) Average F1 score of sucking and swallowing detection over the time. }
    \label{fig:matrix}
\end{figure}

\textbf{Sucking and swallowing detection.}
In total, we annotated \rev{9897} sucking events and \rev{4283} swallowing events across \rev{239} minutes of breastfeeding data. 
For sucking event detection, Mammal achieved a mean F1 score of \rev{0.92 (SD=0.03, P=0.91, R=0.93)}. Swallowing events detected using the garment’s acoustic sensing pipeline achieved a mean F1 score of \rev{0.93 (SD=0.03, P=0.92, R=0.93)} relative to externally monitored ground truth. Fig.~\ref{fig:matrix}a and Fig.~\ref{fig:matrix}b show the confusion matrices for sucking and swallowing event detection, respectively. We do not report true negatives because the data are dominated by non-event windows, which would overwhelm accuracy. Instead, we use F1, which depends on precision and recall and does not require true negatives. 

Since our detection algorithm uses unsupervised clustering that can adapt as more data accumulate, we expect detection performance to change as a feeding session progresses. To quantify this, we computed the F1 score at increasing elapsed times from the start of feeding. Specifically, we evaluated performance from 30\,s to 480\,s (\rev{a common early-session analysis window shared across all sessions}), in 15\,s increments, and aggregated results across sessions (Fig.~\ref{fig:matrix}c).
In turn, we observed a short warm-up period. Within the first 45\,s, F1 scores were lower for both detectors (swallowing: $\sim$0.8; sucking: $\sim$0.7 in the first 30\,s window). After $\sim$45\,s, performance improved and remained stable, with both sucking and swallowing detection converging to $\sim$0.9 for the remainder of the analyzed duration.
This suggests that clustering is a good strategy for quickly adapting decision boundaries to each dyad's signal characteristics, yielding more personalized and reliable detection after a brief initialization phase.

Finally, we examined whether breast size (proxied by bra cup size) affected performance, as it could influence inter-body transmission of acoustic signals. Among the two dyads with cup size C, we did not observe a noticeable difference in detection accuracy compared to the dyads with cup size D (p=0.92). While this result is encouraging, our sample size within each cup-size group was limited. Thus, we treat this analysis as exploratory and leave a larger-scale investigation of body-shape factors to future work.

\textbf{SSB ratio estimation.}
\rev{Based on the detected sucking and swallowing events, Mammal also identifies breathing events from the separated infant ECG and uses these three event streams to estimate the SSB ratio over time. We evaluate ECG-derived breathing events against reference breath events obtained from the infant PPG recording; the two streams are placed on the common session timeline, and a detected breathing event is counted as correct if it falls within 0.3~s of the nearest reference event.} \rev{Compared with ground truth, Mammal achieved F1 score 0.98~(Precision= 0.98, Recall= 0.98) for breathing event detection and a respiration rate MAE of 3.77 breaths/min.} Mammal further estimated the SSB ratio with a mean absolute error of \rev{0.12 (SD = 0.05)} per 10-second window, suggesting that it can capture time-varying feeding coordination during natural breastfeeding.

\begin{figure}[h]
    \centering
    \includegraphics[width=0.6\textwidth]{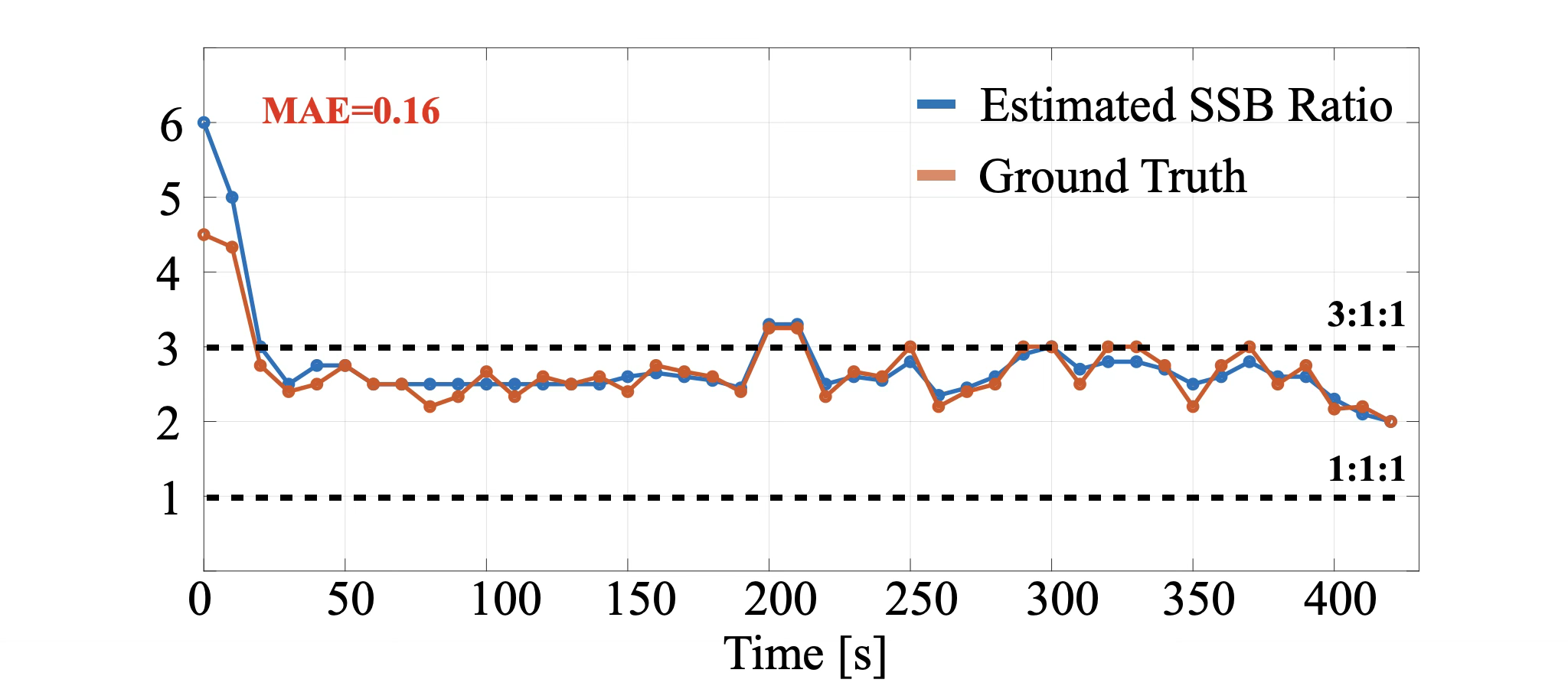}
    \caption{Mammal-estimated SSB ratio aligns with the annotated ground-truth SSB ratio over time.}
    \label{fig:SSBratio}
\end{figure}

Fig.~\ref{fig:SSBratio} shows an example of the estimated and ground-truth SSB ratios over a feeding session from a 2-month-old infant. The ratio is higher and more variable at the beginning of the session, which may reflect a brief period of non-nutritive sucking and latch adjustment. After $\sim$30\,s, the infant settles into a more regular pattern, approaching a consistent 2:1:1 coordination. This stabilization is consistent with effective nutritive feeding, where swallows are regularly interleaved with breathing and sucking bursts, reflecting coordinated airway protection and milk transfer.

\textbf{Milk intake estimation.}
We obtained \rev{9} successful intake reference measurements using pre- and post-feeding infant weights, with observed weight changes ranging from \rev{61--125\,g}. One session lacked a reference measurement because the infant fell asleep mid-feed and the caregiver declined post-feed weighing to avoid waking the infant. 

Overall, milk intake was estimated with a mean relative error of \rev{15.76\%} compared with reference intake from pre/post-feed weighing. This level of accuracy is promising because test weighing is operationally burdensome and, outside tightly controlled protocols (e.g., at home), can show high variability up to 40\% \cite{testweighing_Savenije}. In contrast, Mammal estimates intake passively during natural breastfeeding without requiring caregivers to weigh each feed, enabling day-to-day tracking and aggregation of intake patterns in home settings.

Looking into the per-dyad results, we also observed higher intake errors for two infants younger than 6 months (\rev{23.1\%} and 35.8\%). This pattern may suggest that the mapping between detected sucking intervals and milk transfer is age-dependent. \rev{We therefore developed an age-calibrated model by splitting the cohort into infants younger than 6 months ($N=4$) and infants 6 months or older ($N=5$). For each participant, we applied a leave-one-out evaluation. Specifically, we fit group-specific parameters $a$, $b$, and a maximum interval cap in Eq.~\ref{eq:milk_regression_inverted} using the remaining participants in the same age group, and then evaluated the fitted model on the held-out participant. This age-calibrated model improved accuracy, achieving a MAPE of 4.76\% in the $<6$-month group and 2.91\% in the $\geq 6$-month group. Note that these results are based on small subgroup sizes (N=4 and N=5) and should be interpreted as preliminary evidence. Larger age-stratified datasets are needed to validate the performance gains of the age-calibrated model.}


\begin{figure}[h!]
    \centering
    \includegraphics[width=0.7\linewidth]{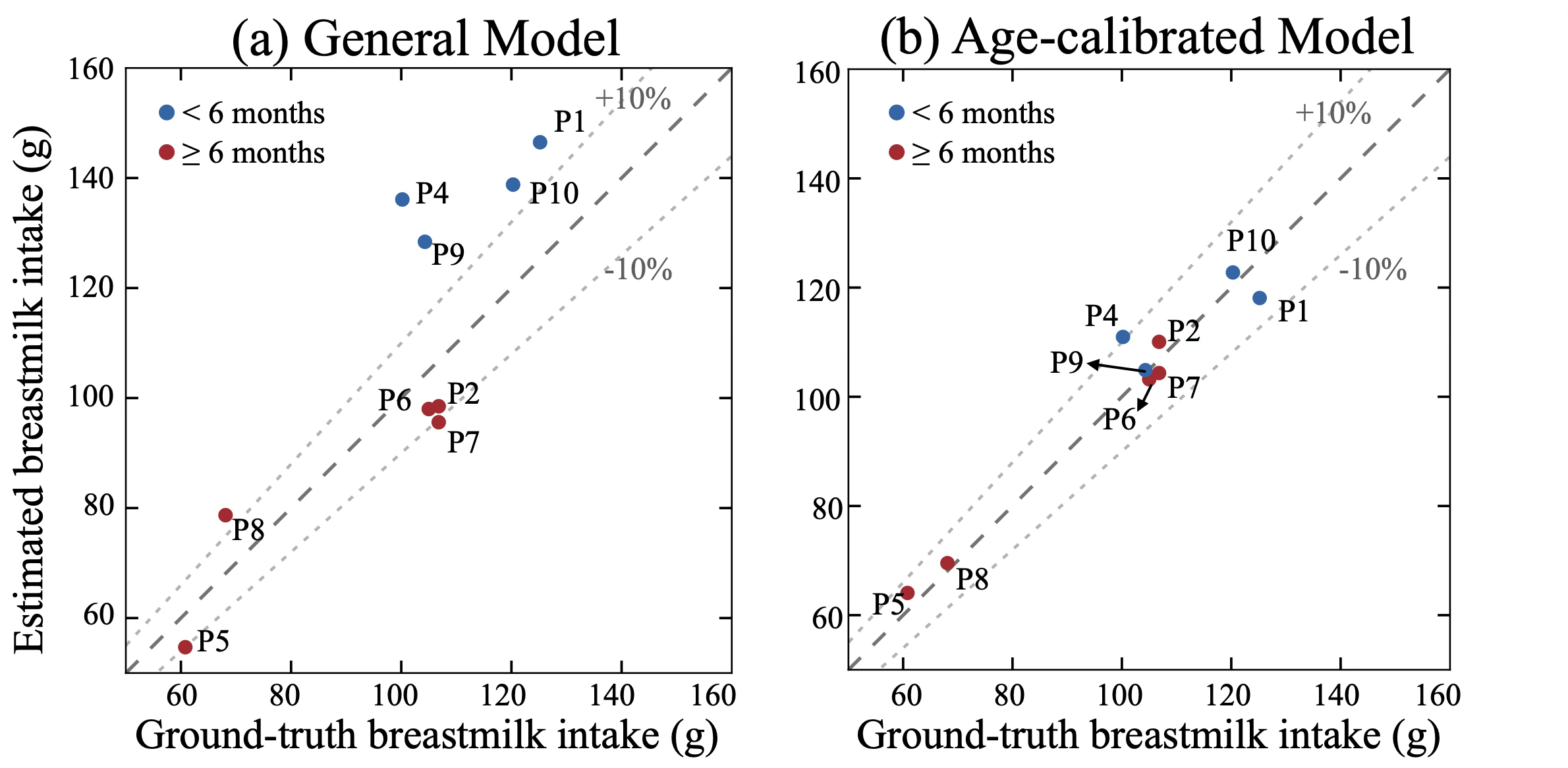}
\caption{
Estimated versus ground-truth breastmilk intake across sessions using 
(a) the original ISI-based milk-intake model and 
(b) the age-calibrated leave-one-out model. 
Each point represents one breastfeeding session with test-weighing ground truth. 
Blue and red points indicate sessions from infants younger than 6 months and 6 months or older, respectively. 
The central dashed line indicates perfect agreement ($y=x$), and the upper and lower dashed lines indicate $\pm$10\% relative error bands.
}
\label{fig:placeholder}
\end{figure}

\subsection{Sensing Micro-benchmarks}\label{Sensing Micro-benchmarks}

\rev{\textbf{ECG Validation.}
To understand Mammal's ECG acquisition quality against a clinical reference, we conducted a preliminary validation study. As placing adhesive ECG electrodes on infant skin is infeasible, we performed this study with three adult participants. For each participant, we simultaneously recorded ECG during a 20-minute session using Mammal and a clinical-grade ECG device (Biocare IE300~\cite{biocare_ie300}), with both systems configured in the same lead-I setup: one electrode on each arm and an additional reference electrode placed on the chest as a common reference. This setup enables a controlled comparison between Mammal and the clinical reference under matched conditions. Although this protocol may not reproduce the full caregiver--infant breastfeeding scenario, it allows us to assess whether Mammal's textile-electrode acquisition pipeline can capture cardiac waveforms and support accurate heart-rate estimation.}

\begin{figure}[h]
    \centering

    \begin{minipage}[t][4.5cm][b]{0.36\linewidth}
        \centering
        \includegraphics[width=0.95\linewidth]{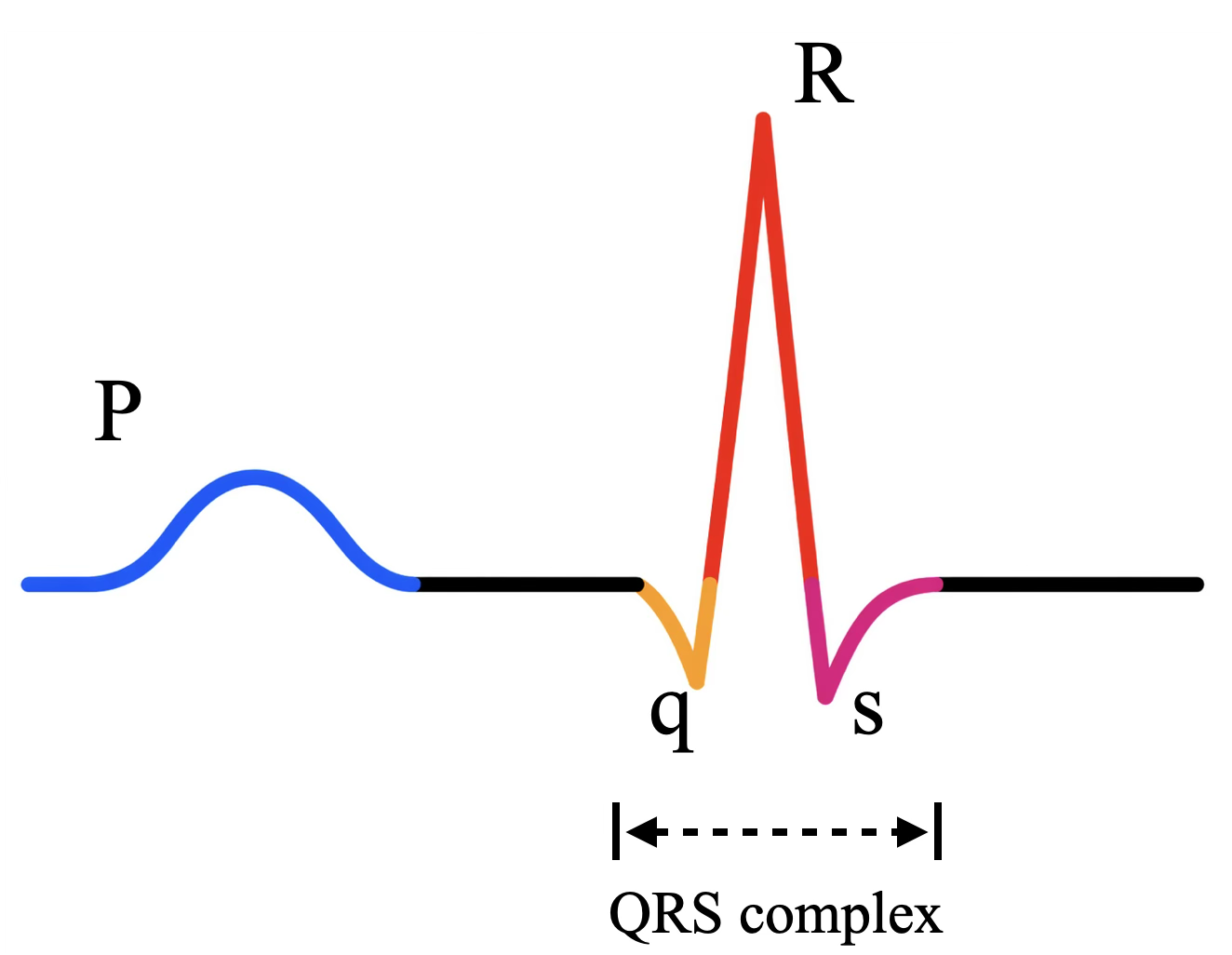}

        \vfill
        \textbf{(a)} The QRS complex in ECG waveform.
    \end{minipage}
    \hfill
    \begin{minipage}[t][4.1cm][b]{0.58\linewidth}
        \centering
        \setlength{\tabcolsep}{4pt}
        \renewcommand{\arraystretch}{1.05}
        \begin{tabular}{lccc}
            \toprule
            \textbf{Participant} & \textbf{Duration} & \textbf{QRS Corr.} & \textbf{HR MAE} \\
            & \textbf{(min)} & \textbf{($r$)} & \textbf{(bpm)} \\
            \midrule
            P1 & 20 & 0.95 & 0.22 \\
            P2 & 20 & 0.95 & 0.62 \\
            P3 & 20 & 0.97 & 0.19 \\
            \midrule
            Mean $\pm$ SD & 60 in total & 0.96 $\pm$ 0.01 & 0.34 $\pm$ 0.20 \\
            \bottomrule
        \end{tabular}

        \vfill
        \textbf{(b)} Clinical-grade ECG validation results.
    \end{minipage}

    \caption{
    Clinical-grade ECG validation across three adult participants. 
    (a) The ECG waveform highlights the QRS complex used for QRS-centered waveform correlation. 
    (b) Quantitative validation results report QRS-centered waveform correlation and heart-rate estimation error using the Biocare IE300 medical-grade ECG reference.
    }
    \label{fig:clinical_ecg_validation}
\end{figure}

\rev{We report two metrics: QRS-centered waveform correlation and heart-rate estimation error. The QRS complex is the most prominent portion of the ECG waveform and corresponds to ventricular depolarization, making it a landmark for comparing cardiac waveform morphology across system (Fig.~\ref{fig:clinical_ecg_validation}a). Because Mammal focuses on heart-rate and respiratory-rate estimation rather than diagnostic ECG interpretation, the QRS complex and R-peak timing provide sufficient information for downstream estimation~\cite{dogan2023comprehensive, 10.1371/journal.pdig.0000538, roberts2024opensource}. After temporally aligning the ECG signals recorded by the two systems, we compute the local Pearson correlation within QRS-centered windows to quantify waveform similarity. Specifically, for each ground-truth R peak, we extracted paired 250~ms windows spanning $\pm125$~ms from the separated and ground-truth ECG signals, computed their Pearson correlation, and averaged the correlations across all detected R peaks~\cite{martins2025textile_ecg_quality}. We also derive heart rate from the detected R-peaks in each signal and report the MAE in beats per minute. Fig.~\ref{fig:clinical_ecg_validation}b summarizes the results across the three adult participants. Mammal achieved a mean QRS-centered waveform correlation of 0.96 (SD= 0.01) and a mean heart-rate MAE of 0.34 bpm (SD= 0.20) relative to the Biocare IE300 reference. These results suggest that Mammal's textile-electrode acquisition pipeline can provide ECG measurements of sufficient quality for downstream heart-rate estimation, aligning with findings in prior research \cite{tsukada2019validation,shao2024joey,s20216233}.}

\rev{\textbf{ECG Separation During Heartbeats Overlap.} Theoretically, overlapping heartbeats could make infant ECG separation more challenging because the adaptive filter may overfit the caregiver ECG component in the inter-body ECG when the two cardiac events occur nearly simultaneously. However, such overlap is relatively uncommon because adult and infant heart rates typically differ substantially (about 60--100~bpm for adults and 100--160~bpm for infants). To quantify the robustness of our pipeline under this condition, we analyzed recordings from four pilot adult dyads with ground-truth ECG available for both individuals. We defined an overlap beat as a pair of R-peaks occurring within 50~ms of each other across the two individuals. This threshold approximates the temporal interval over which two QRS complexes would morphologically interfere in the inter-body signal, based on prior reports of typical adult R-wave peak time \cite{doi:10.1161/CIRCULATIONAHA.108.191095,pava2010rwave}. Based on this criterion, we identified 249 overlapping beats out of 2966 total detected beats (8.4\%). We then compared the separated ECG against ground truth using QRS-centered Pearson correlation and beat-wise instantaneous HR error. Mammal achieved a mean QRS-centered Pearson correlation of 0.94 for non-overlapping beats and 0.93 for overlapping beats, and an HR MAE of 1.72 bpm and 2.18 bpm, respectively. A representative qualitative example is shown in Appendix~\ref{appendix_overlapping}.
These results indicate that heartbeat overlap causes only modest performance degradation, while the proposed ECG separation pipeline still preserves local waveform morphology and supports reliable heart-rate estimation.}

\rev{\textbf{Ablation Study on Temporal Segmentation, DWT Decomposition Level, and Conditional Denoising.}\label{segmentation}
To examine the effects of temporal segmentation, DWT decomposition depth, and conditional denoising in the ECG separation pipeline, we conducted ablation analyses on both the 4-dyad adult pilot dataset and 10 caregiver--infant dyad dataset. For the adult pilot dataset, where waveform-level ground-truth ECG was available, we evaluated performance using mean QRS-centered local waveform correlation~\cite{martins2025textile_ecg_quality}. For the caregiver--infant dyad dataset, where clean infant ECG waveforms were unavailable, we evaluated performance using infant heart-rate MAE.}

\rev{ Table~\ref{tab:ablation_pipeline} summarizes the results. Temporal segmentation consistently improved performance over the non-segmented setting, increasing waveform correlation from 0.87 (SD=0.03) to 0.94 (SD=0.03) and reducing HR MAE from 5.12 bpm (SD=1.67) to 3.61 bpm (SD= 0.87). We further compared two temporal segmentation settings, using either two 0.5~s segments or three 0.33~s segments. While the three-segment setting maintained a similar HR MAE, it reduced the QRS-centered local waveform correlation compared with the two-segment setting. This result suggests that finer segmentation may help suppress residual noise for heart-rate estimation, but at the cost of disrupting local waveform morphology.
Under the temporally segmented two-segment setting, DWT Levels 2 and 3 yielded similar performance, while Level 4 achieved the best overall results. In contrast, performance degraded at Level 5, suggesting that deeper decomposition does not necessarily improve separation quality.}


\begin{table}[h]
\centering
\caption{Ablation results for temporal segmentation, DWT decomposition level, and conditional denoising in the ECG separation pipeline. For the adult pilot dataset, performance is evaluated using mean QRS-centered local waveform correlation. For the caregiver--infant dyad dataset, performance is evaluated using infant heart-rate MAE. Higher correlation and lower MAE indicate better performance. Denoising is evaluated on triggered 1~s windows.}
\label{tab:ablation_pipeline}
\setlength{\tabcolsep}{1.5pt}
\begin{tabular}{llcc}
\toprule
\textbf{Ablation factor} & \textbf{Setting} & \textbf{Adult dyads} & \textbf{Caregiver--infant dyads} \\
 &  & \textbf{QRS local corr.} & \textbf{HR MAE (bpm)} \\
\midrule
Temporal segmentation & Without segmentation & 0.87 & 5.12 \\
Temporal segmentation & With segmentation (two 0.5~s segments) & 0.94 & 3.61 \\
Temporal segmentation & With segmentation (three $\sim$0.33~s segments) & 0.85 & 3.47 \\
\midrule
DWT decomposition depth & Level 2 & 0.93 & 3.75 \\
DWT decomposition depth & Level 3 & 0.92 & 3.67 \\
DWT decomposition depth & Level 4 & 0.94 & 3.61 \\
DWT decomposition depth & Level 5 & 0.89 & 5.84 \\
\midrule
Conditional denoising & Butterworth only & 0.93 & 5.80 \\
Conditional denoising & Butterworth + DeScoD-ECG & 0.91 & 3.61 \\
\bottomrule
\end{tabular}
\end{table}

\rev{
We further evaluated the contribution of conditional DeScoD-ECG denoising on the subset of 1~s windows for which the denoising trigger fired. Because non-triggered windows are processed identically by both pipelines, they were excluded from this comparison. On those valid windows, Butterworth-only processing yielded an HR MAE of 5.80 bpm (SD= 1.85), whereas Butterworth followed by DeScoD-ECG reduced the error to 3.61 bpm (SD= 0.87), corresponding to a 37.8\% reduction. For the QRS-centered local waveform correlation, we found that Butterworth-only processing preserve more waveform, indicating that DeScoD-ECG, though reducing the suspect peak caused motion artifact, also compromises the waveform. An example waveform illustrating this conditional denoising behavior is provided in Appendix~\ref{appendix_descod}. Together, these results indicate that temporal segmentation improves local waveform preservation, Level-4 DWT provides the best trade-off between representation richness and separation robustness, and conditional denoising further removes residual in-band contamination when severe motion-induced artifacts remain.}


\rev{\textbf{Leave-one-dyad-out Sucking and Swallowing Detection.}
Our study results show that the feature-based fuzzy clustering pipeline can adaptively and unsupervisedly detect sucking and swallowing based on 20 labeled initialization samples from a single pilot dyad, with F1 scores of 0.92 (SD=0.03, P=0.91, R=0.93) and 0.93 (SD=0.03, P=0.92, R=0.93), respectively. This finding is encouraging because the pipeline was initialized using only 20 labeled samples from a single pilot dyad. To further investigate whether detection performance could improve with more participant data, we conducted a leave-one-dyad-out (LODO) analysis across the ten caregiver--infant dyads.
In each fold, one dyad was held out for testing, while labeled initialization samples drawn from the remaining nine dyads were used to initialize the fuzzy clustering model. As a result, the LODO initialization achieved a mean F1 of 0.94 (SD=0.02, P=0.94, R=0.94) for sucking detection and 0.93 (SD=0.02, P=0.93, R=0.93) for swallowing detection.  This result suggests that the sensing pipeline is already robust with limited initialization data, while additional participant data only slightly improve sucking detection performance.}

 \subsection{User Experience}
\begin{figure}[h]
   \centering
   \includegraphics[width=0.4\linewidth]{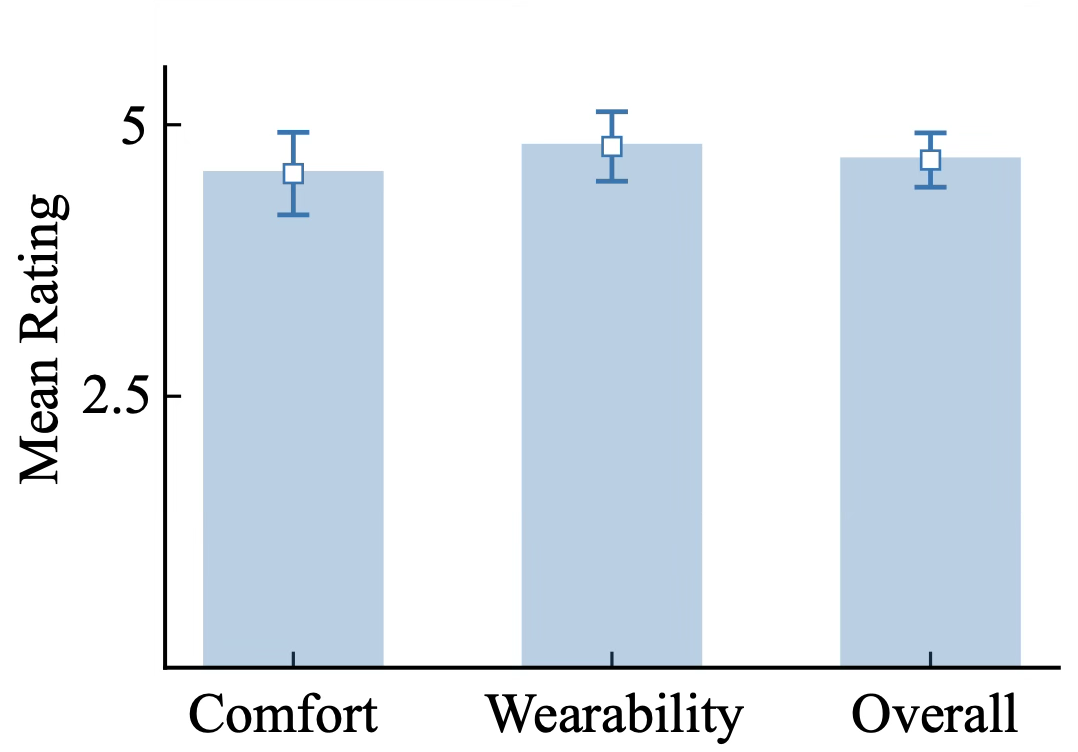}
   \caption{Mean caregiver ratings of the Mammal garment on comfort, wearability, and overall satisfaction.}
   \label{fig:score}
 \end{figure}

 We assessed caregiver user experience along three dimensions: comfort, wearability, and usability, using post-study questionnaires and a semi-structured interview conducted after the breastfeeding session.  Fig.~\ref{fig:score} summarizes the questionnaire ratings.

 \textbf{Comfort.}
 Caregivers reported overall comfort while wearing the Mammal garment during breastfeeding, with a mean comfort rating of \rev{4.62 out of 5 (SD = 0.28)}. \rev{Breastfeeding postures observed during the study were primarily cradle, cross-cradle, and football hold.} Most caregivers noted that the garment did not interfere with infant positioning or skin-to-skin contact. A small number of participants reported minor discomfort related to strap tension, electrode pressure, which was typically resolved through garment adjustment.

 \textbf{Wearability.}
 Wearability was evaluated in terms of fit, stability, and freedom of movement during feeding. Caregivers rated the garment as \rev{4.75 out of 5} for wearability \rev{(SD = 0.19)}, and reported that it remained stable across common breastfeeding postures, including seated and reclined positions. Participants emphasized that the garment allowed natural movement and did not restrict routine caregiving activities.

 \textbf{User Feedback.}
 We collected post-study feedback focusing on comfort, privacy, and perceived safety. Overall, caregivers reported low privacy concerns because the system is worn only during breastfeeding sessions. Participants also noted that the double-clip design improved nursing access by enabling quicker and easier latch and adjustment. However, two caregivers reported noticing the contact microphone while it was placed on the lower band of the bra, even when covered by a cotton pad. One participant also pointed out that the back-mounted enclosure felt bulky and could draw the infant’s attention during feeding, with some infants attempting to touch or press against the box.


\section{DISCUSSION AND FUTURE WORK}

\subsection{Preliminary Expert Review.}
To understand Mammal's clinical relevance and identify practical applications in neonatology, we conducted a preliminary interview with a neonatologist working in a local Neonatal Intensive Care Unit (NICU). The interview focused on three topics: (1) potential end users, (2) likely usage scenarios, and (3) pathways for integration into clinical workflows. 

Overall, the neonatologist viewed Mammal as clinically valuable due to its detection of SSB coordination and heart rate. He highlighted that NICU-to-home transition infants are strong candidates because physiologic vulnerability can persist after discharge while observation drops sharply. In particular, term/near-term infants with comorbidities (e.g., chronic lung disease, congenital heart disease, neurologic injury) may benefit from monitoring of suck--swallow--breathe coordination and autonomic stability, while preterm infants may require more accurate intake estimation to guide breastfeeding suitability and supplementation planning.
The clinician also emphasized potential applications in risk-aware timely alerts and trend summaries, such as warning coordination abnormalities, prolonged pauses, or concurrent heart-rate deviations that could precede clinically relevant events (e.g., bradycardia, choking and silent aspiration). He noted this is especially important during night feeds or caregiver fatigue, when attention to infant feeding cues is reduced.
For workflow integration, the neonatologist recommended embedding Mammal into discharge readiness and follow-up, such as introducing it during the final NICU/step-down phase for caregiver training, then use post-discharge summaries and longitudinal trends to support remote monitoring and targeted follow-ups (e.g., lactation consults, feeding therapy, pediatric visits).

\subsection{Failure Mode and Practical Limitation}

\rev{A key limitation of Mammal is that inter-body ECG relies on physical contact between the infant's body, such as the hand, and the garment's waist electrode to establish the electrical pathway. Because of this, Mammal cannot capture inter-body ECG signals when the infant is swaddled, heavily clothed, or positioned such that no direct contact is made with the waist electrode. This limitation may be especially relevant in colder environments or outdoor settings, where infants are more likely to be covered. A possible future direction is to explore textile-based non-contact capacitive electrodes, combined with actively shielded readout electronics, to reduce reliance on direct skin contact \cite{https://doi.org/10.1155/2021/6698567,CHEN2025e00718}.}

Another practical limitation is that our system was evaluated in a relatively controlled setting. In-home deployment will introduce more dynamic factors that were not fully captured in our sessions, including environmental noise, frequent infant repositioning, unstable latch transitions, and caregiver movement. Although Mammal’s acoustic sensing is less directly affected by airborne noise due to its use of contact microphones, these real-world factors may still introduce motion artifacts, perturb inter-body coupling, and affect event detection. Future work will develop noise- and artifact-aware acoustic processing methods to improve robustness in more natural breastfeeding environments.

Finally, our current system operates offline, with signals processed after collection rather than in real time. Therefore, the present prototype does not yet support live monitoring or immediate feedback during feeding. In future work, we plan to develop a real-time implementation with online signal processing, signal-quality assessment, and artifact handling to improve practicality in everyday breastfeeding scenarios.


\subsection{Study Limitation and Longitudinal Validation}
Our user study is limited in sample size ($N=10$) and participant age (2--15 months). One practical reason is that the study was conducted in a private room within a clinical research and trial unit, requiring participants to coordinate breastfeeding timing and travel to the site. Given the small sample size, this work represents a feasibility demonstration of Mammal rather than a rigorous validation for clinical deployment. While the findings are encouraging, larger-scale studies are needed to establish robustness, generalizability, and clinical utility across diverse caregivers, infants, and feeding conditions. A larger cohort would also enable further development and validation of the age-dependent milk intake estimation model.
In future work, we plan to develop a real-time system and a study protocol that would allow participants to take part from home. Despite introducing more noise, in-home deployment would make participation more practical for breastfeeding dyads and reduce the burden of travel and scheduling, enabling a longitudinal study with a larger cohort and repeated sessions per dyad. Longitudinal data would allow us to further strengthen the evidence base by characterizing day-to-day variability, quantifying within-dyad reliability and between-dyad generalization, and evaluating whether sensing performance and usability remain robust over time. It would also support refinement of the milk intake model using a broader and more representative dataset, reducing the risk of overfitting and improving robustness across developmental stages.


\subsection{Washability and Maintenance}
Washability and maintenance are critical for caregiver-worn systems deployed in everyday feeding. While we do not directly solve this issue, a practical direction is to make rigid electronics detachable and to protect sensitive components (e.g., microphones) via sealing, encapsulation, or modular replacement. Prior work on textile sensing has shown that conductive fabrics and textile electrodes can be designed to tolerate repeated washing and drying under appropriate construction and care protocols. Nonetheless, microphone sealing and cable strain relief remain challenges for repeated laundering. Future iterations of Mammal will prioritize a fully detachable electronics module, wash-safe textile routing, and standardized care instructions (e.g., gentle cycle, air drying), followed by durability testing across repeated wash cycles to quantify changes in signal quality and mechanical integrity.

\section{Conclusion}
We present Mammal, a computational garment that leverages inter-body signal transmission to unobtrusively monitor breastfeeding physiology, including latch onset, duration, infant heart rate, suck–swallow–breathe ratio, and milk intake, without instrumenting the infant. In a study with \rev{10} caregiver–infant dyads, Mammal achieved a \rev{5.56\%} mean absolute percentage error for latch duration, a \rev{3.61} bpm mean absolute error for infant heart rate, a \rev{0.12} MAE for SSB ratio, and a \rev{15.76\%} mean relative error for milk intake, while participants reported high comfort and wearability. A preliminary NICU clinician interview  highlighted Mammal’s promise for supporting the NICU-to-home transition through caregiver-worn monitoring during natural breastfeeding.
Future work will address washability and maintenance to enable long-term deployment, and will validate Mammal through larger, longitudinal in-home studies across diverse dyads and conditions. More broadly, by enabling the caregiver body as a sensing interface, Mammal opens opportunities to study and ultimately support maternal–infant well-being and co-regulation in everyday settings.

\begin{acks}

We thank our reviewers for their constructive and insightful comments. We also thank all participants in our studies, including the adult participants and caregiver--infant dyads, for their time and valuable feedback on the Mammal system. We thank the Florida State University Clinical Research and Trials Unit for their support during participant recruitment and study sessions. We thank the funding support from the Florida Institute for Pediatric Rare Diseases for developing this project.

\end{acks}




\bibliographystyle{ACM-Reference-Format}
\bibliography{sample-base}

\appendix

\section{Additional Study Details}
\label{sec:appendix_details}

\subsection{Participant Demographics}\label{demographics}
Table~\ref{tab:participants} summarizes the demographic information of the caregiver--infant dyads in our breastfeeding study, including caregiver age, infant age and sex, as well as the bra cup size and garment size used during each session.
\begin{table}[h!]
  \centering
  \caption{Participant demographics (10 caregiver--infant dyads).}
  \label{tab:participants}
  \vspace{-0.5em}
  \begin{tabular}{lccccc}
    \toprule
    Dyad & Caregiver age (years) & Infant age (months) & Infant sex & Garment size & Bra cup size \\
    \midrule
    P1 & 31 & 3 & Male   & Medium & 36C  \\
    P2 & 37 & 8 & Female & Medium & 34DD \\
    P3 & 33 & 11 & Male   & Small  & 32C  \\
    P4 & 39 & 2  & Male   & Large  & 34D  \\
    P5 & 26 & 12 & Male   & Large  & 36DD \\
    P6 & 28 & 11  & Male   & Small  & 34D  \\
    P7 & 32 & 15  & Female & Small  & 34DD \\
    P8 & 34 & 9  & Female & Large  & 36D  \\
    P9 & 34 & 4 & Female   & Large  & 44DDD  \\
    P10 &28 & 3  & Male & Large  & 36DD  \\
    \bottomrule
  \end{tabular}

\end{table}

\begin{figure}[h!]
  \centering
  \includegraphics[width=0.55\linewidth]{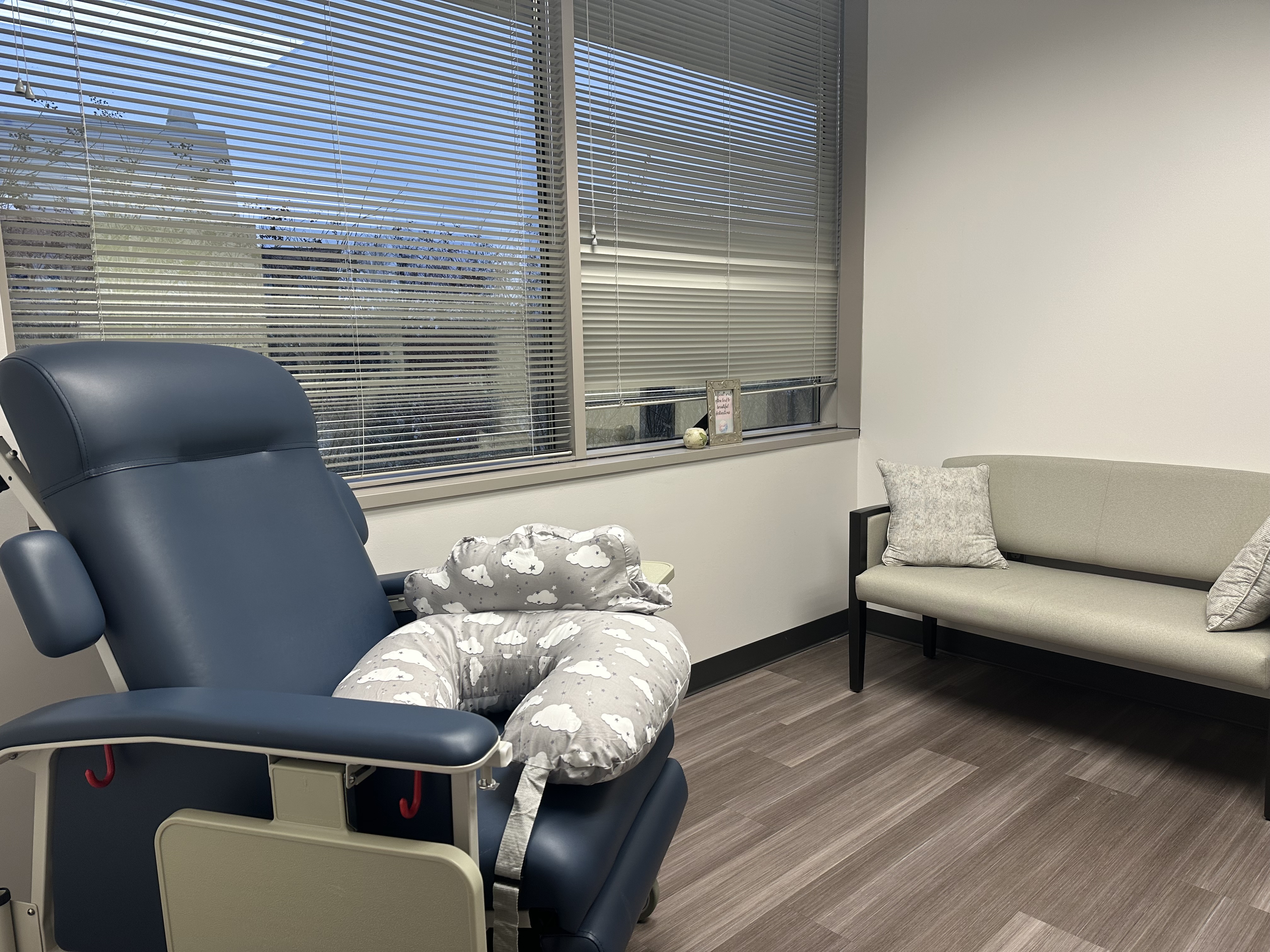}
  \caption{Study environment used in the breastfeeding sessions. The room included a breastfeeding couch and a long chair to support different feeding postures.}
  \label{fig:room}
\end{figure}

\subsection{Study Environment}\label{studyroom}

Fig.~\ref{fig:room} shows the study room setup used for the breastfeeding sessions, including a breastfeeding couch and a long chair to accommodate different feeding postures.


\subsection{Detailed Per-Session Results}\label{appendix_per_session_results}
\rev{This section provides detailed results for all 10 analyzed breastfeeding sessions. The following tables report session-level results for latch onset and duration estimation, infant physiological estimation, and sucking/swallowing detection.}

\rev{\textbf{Latch Onset Detection and Latch Duration Estimation.}
Table~\ref{tab:appendix_latch_detail} reports the detailed per-session results for latch onset detection and latch duration estimation, including event-level detection performance and duration error.}

\begin{table*}[h!]
\centering
\caption{Per-session latch onset detection and latch duration estimation results.}
\label{tab:appendix_latch_detail}
\setlength{\tabcolsep}{4pt}
\begin{tabular}{lcccccc}
\toprule
\textbf{Session} &
\textbf{GT onsets} &
\textbf{TP / FP / FN} &
\textbf{P / R / F1} &
\textbf{GT latch} &
\textbf{Est. latch} &
\textbf{Duration} \\
& & & &
\textbf{duration (min)} &
\textbf{duration (min)} &
\textbf{error(\%)} \\
\midrule
P1  & 13 & 13 / 3 / 0  & 0.81 / 1.00 / 0.90 & 31.2 & 29.3 & 6.09 \\
P2  & 11 & 11 / 1 / 0  & 0.92 / 1.00 / 0.96 & 26.3 & 27.3 & 3.80 \\
P3  & 9  & 9 / 0 / 0   & 1.00 / 1.00 / 1.00 & 18.9 & 21.3 & 12.70 \\
P4  & 8  & 8 / 0 / 0   & 1.00 / 1.00 / 1.00 & 23.9 & 25.8 & 7.95 \\
P5  & 4  & 3 / 0 / 1   & 1.00 / 0.75 / 0.86 & 20.2 & 21.5 & 6.44 \\
P6  & 10 & 9 / 0 / 1   & 1.00 / 0.90 / 0.95 & 15.9 & 15.2 & 4.40 \\
P7  & 11 & 10 / 1 / 1  & 0.91 / 0.91 / 0.91 & 28.2 & 27.4 & 2.84 \\
P8  & 8  & 7 / 1 / 1   & 0.88 / 0.88 / 0.88 & 17.4 & 16.9 & 2.87 \\
P9  & 16 & 15 / 3 / 1  & 0.83 / 0.94 / 0.88 & 24.7 & 23.9 & 3.24 \\
P10 & 12 & 11 / 1 / 1  & 0.92 / 0.92 / 0.92 & 32.5 & 34.2 & 5.23 \\
\midrule
Mean $\pm$ SD & - & - &
0.93$\pm$0.07 / 0.93$\pm$0.08 / 0.93$\pm$0.05
& - & - & 5.56$\pm$2.87 \\
\bottomrule
\end{tabular}
\end{table*}


\rev{\textbf{Infant Heart Rate, Respiration Rate, and SSB Ratio Estimation.}
Table~\ref{tab:appendix_continuous_detail} summarizes the per-session continuous-estimation results for infant heart rate, respiration rate, and SSB ratio.}

\begin{table*}[h!]
\centering
\caption{Per-session continuous-estimation results for infant heart rate, respiration rate, and SSB ratio.}
\label{tab:appendix_continuous_detail}
\setlength{\tabcolsep}{6pt}
\begin{tabular}{lccc}
\toprule
\textbf{Session} & 
\textbf{HR MAE} & 
\textbf{RR MAE} & 
\textbf{SSB MAE} \\
& 
\textbf{(bpm)} & 
\textbf{(breaths/min)} & 
\textbf{(per session)} \\
\midrule
P1  & 2.6 & 4.2 & 0.20 \\
P2  & 3.2 & 2.1 & 0.08 \\
P3  & 3.4 & 4.5 & 0.13 \\
P4  & 5.3 & 4.7 & 0.16 \\
P5  & 2.4 & 2.9 & 0.18 \\
P6  & 2.7 & 4.3 & 0.09 \\
P7  & 4.1 & 3.6 & 0.06 \\
P8  & 4.3 & 3.8 & 0.05 \\
P9  & 3.9 & 3.1 & 0.11 \\
P10 & 4.2 & 4.5 & 0.15 \\
\midrule
Mean $\pm$ SD & 3.61$\pm$0.87 & 3.77$\pm$0.80 & 0.12$\pm$0.05 \\
\bottomrule
\end{tabular}
\end{table*}

\rev{\textbf{Sucking and Swallowing Detection.} Table~\ref{tab:appendix_suck_swallow_detail} presents the per-session event-detection results for sucking and swallowing, including the number of ground-truth events, TP/FP/FN counts, and the derived precision, recall, and F1 scores.}
\begin{table*}[h!]
\centering
\caption{Per-session sucking and swallowing detection results.
P / R / F1 in the last row are reported as macro-averaged 
mean~$\pm$~SD across the 10 sessions, using population SD.}
\label{tab:appendix_suck_swallow_detail}
\small
\setlength{\tabcolsep}{3.5pt}
\newcommand{\msd}[2]{$#1\pm #2$}
\begin{tabular}{@{}lcccccc@{}}
\toprule
\textbf{Session} &
\multicolumn{3}{c}{\textbf{Sucking}} &
\multicolumn{3}{c}{\textbf{Swallowing}} \\
\cmidrule(lr){2-4} \cmidrule(lr){5-7}
& \textbf{$N_{\mathrm{GT}}$} & \textbf{TP / FP / FN} & \textbf{P / R / F1}
& \textbf{$N_{\mathrm{GT}}$} & \textbf{TP / FP / FN} & \textbf{P / R / F1} \\
\midrule
P1  & 1576 & 1392 / 289 / 184 & 0.83 / 0.88 / 0.85 & 616 & 528 / 87 / 88 & 0.86 / 0.86 / 0.86 \\
P2  &  889 &  842 /  37 /  47 & 0.96 / 0.95 / 0.95 & 407 & 384 / 21 / 23 & 0.95 / 0.94 / 0.95 \\
P3  & 1137 & 1075 /  96 /  62 & 0.92 / 0.95 / 0.93 & 548 & 493 / 57 / 55 & 0.90 / 0.90 / 0.90 \\
P4  &  847 &  809 /  84 /  38 & 0.91 / 0.96 / 0.93 & 348 & 324 / 30 / 24 & 0.92 / 0.93 / 0.92 \\
P5  &  425 &  410 /  33 /  15 & 0.93 / 0.96 / 0.94 & 145 & 136 /  7 /  9 & 0.95 / 0.94 / 0.94 \\
P6  &  875 &  783 /  42 /  92 & 0.95 / 0.89 / 0.92 & 337 & 325 / 29 / 12 & 0.92 / 0.96 / 0.94 \\
P7  & 1127 & 1046 /  85 /  81 & 0.92 / 0.93 / 0.93 & 485 & 441 / 37 / 44 & 0.92 / 0.91 / 0.92 \\
P8  &  696 &  623 /  63 /  73 & 0.91 / 0.90 / 0.90 & 444 & 421 / 25 / 23 & 0.94 / 0.95 / 0.95 \\
P9  & 1019 &  966 / 107 /  53 & 0.90 / 0.95 / 0.92 & 415 & 396 / 22 / 19 & 0.95 / 0.95 / 0.95 \\
P10 & 1306 & 1273 / 128 /  33 & 0.91 / 0.97 / 0.94 & 538 & 512 / 33 / 26 & 0.94 / 0.95 / 0.95 \\
\midrule
Total
& 9897 & 9219 / 964 / 678 & --
& 4283 & 3960 / 348 / 323 & -- \\
Macro avg
& -- & -- & \msd{0.91}{0.03}\,/\,\msd{0.93}{0.03}\,/\,\msd{0.92}{0.03}
& -- & -- & \msd{0.92}{0.03}\,/\,\msd{0.93}{0.03}\,/\,\msd{0.93}{0.03} \\
\bottomrule
\end{tabular}
\end{table*}

\section{Additional Examples for ECG Separation}
\subsection{ECG Separation During Overlapping Heartbeats}\label{appendix_overlapping}
\begin{figure}[h!]
    \centering
    \includegraphics[width=0.8\linewidth]{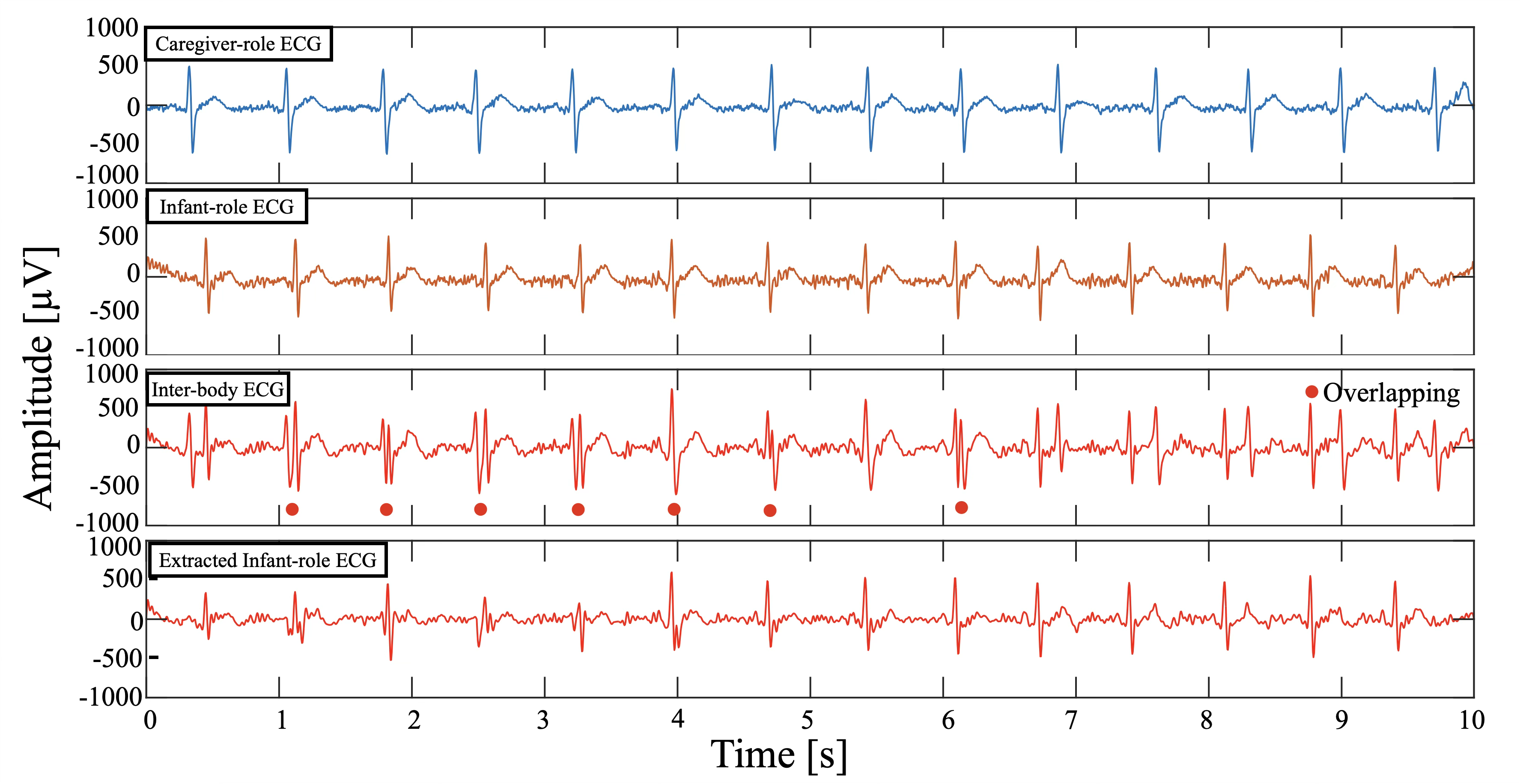}
    \caption{Representative example of ECG separation during temporally     
    overlapping heartbeats in the adult-pair pilot study. 
From top to bottom, the plots show the caregiver-role ECG, infant-role ECG, mixed inter-body ECG, and extracted infant-role ECG. 
Red markers indicate time points where the caregiver-role and infant-role heartbeats overlap in the mixed inter-body ECG.}    \label{fig:overlapping}
\end{figure}

\rev{Fig.~\ref{fig:overlapping} shows a representative example from the adult-pair pilot study, where ground-truth ECG from both participants was available. This example illustrates ECG separation when the two participants' heartbeats temporally overlap.}

\subsection{Conditional DeScoD-ECG Denoising}\label{appendix_descod}

\begin{figure}[h!]
    \centering
    \includegraphics[width=0.9\linewidth]{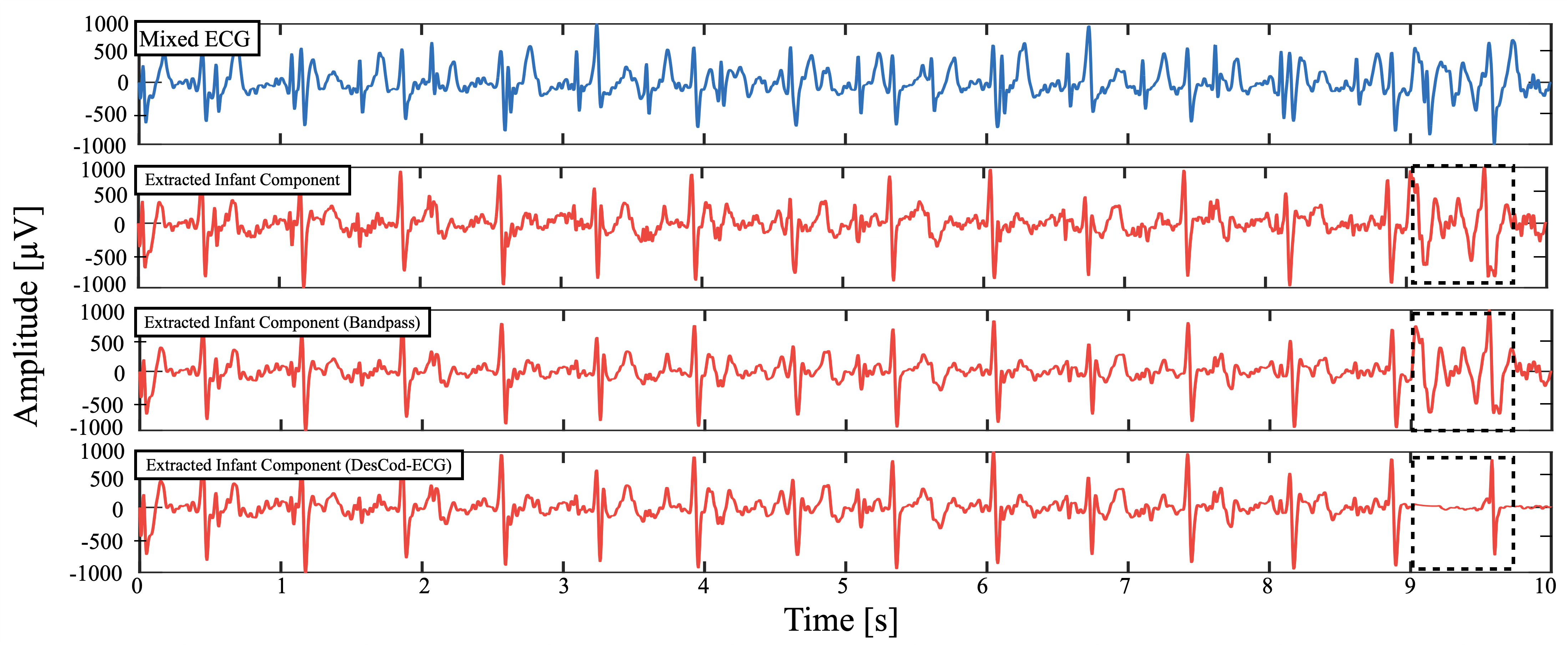}
\caption{Example of conditional DeScoD-ECG denoising. 
From top to bottom, the plots show the mixed inter-body ECG, the extracted infant component before final filtering, the extracted infant component after bandpass filtering, and the output after DeScoD-ECG denoising. 
The dashed boxes highlight a triggered high-artifact segment, where DeScoD-ECG suppresses residual fluctuations while preserving the main heartbeat structure.}
\label{fig:descod}
\end{figure}

\rev{Fig.~\ref{fig:descod} shows a representative example of conditional DeScoD-ECG denoising applied to the extracted infant ECG component.}

\end{document}